\newcommand{\feh}{$[Fe/H]\,$}        
\newcommand{\logg}{$\log g\,$}        
\newcommand{\Flam}{$F_\lambda\,$}        
\newcommand{\Teff}{$T_{\rm eff}\,$}        
\newcommand{\DYDZ}{$\Delta$$Y$/$\Delta$$Z\,$}        
\newcommand{\DM}{$\Delta M\,$}        
\newcommand{\Zsun}{$Z_\odot\,$}        
\newcommand{\Msun}{$M_\odot\,$}        
\newcommand{\Lsun}{$L_\odot\,$}        
\newcommand{\Menv}{$M_{env}\,$}        
\newcommand{\Mcore}{$M_{core}\,$}        
\newcommand{\MRG}{$M_{RG}\,$}        
\newcommand{\MMHB}{$\overline{M_{HB}}\,$} 
\newcommand{\MMHBeq}{\overline{M_{HB}}\,} 
\documentstyle[11pt,aaspp4,psfig]{article}
\slugcomment{ApJ, 486, in press}

\lefthead{Yi et al.}
\righthead{UV Upturn: Sensitivity of UV Population Synthesis}

\begin{document}

\title{On the Origin of the UV Upturn in Elliptical Galaxies. I. Sensitivity of UV Population Synthesis to Various Input Parameters}
\author{Sukyoung Yi\altaffilmark{1}}
\affil{Department of Astronomy, Yale University, P.O. Box 208101, New Haven, CT 06520-8101, NASA/Goddard Space Flight Center, Code 681, Greenbelt, MD 20771\\ email: yi@shemesh.gsfc.nasa.gov}
 
\author{Pierre Demarque and Augustus Oemler, Jr.\altaffilmark{2}}
\affil{Department of Astronomy, Yale University, P.O. Box 208101,
    New Haven, CT 06520-8101 \\ demarque@astro.yale.edu oemler@ociw.edu}

\altaffiltext{1}{National Research Council Research Associate.}
\altaffiltext{2}{Also at Carnegie Observatories, 813 Santa Barbara St., Pasadena, CA 91101}

\begin{abstract}

   We present models of the late stages of stellar evolution intended to
explain the ``UV upturn'' phenomenon in elliptical galaxies. Such models
are sensitive to values of a number of poorly-constrained physical
parameters, including metallicity, age, stellar mass loss, helium
enrichment, and the distribution of stars on the zero age horizontal
branch (HB). We explore the sensitivity of the results to values of
these parameters, and reach the following conclusions.

   Old, metal rich galaxies, such as giant ellipticals, naturally develop
a UV upturn within a reasonable time scale - less than a Hubble time -
without the presence of young stars. The most likely stars to dominate
the UV flux of such populations are low mass, core helium burning (HB
and evolved HB) stars. Metal-poor populations produce a higher ratio
of UV-to-$V$ flux, due to opacity effects, but only metal-rich
stars develop a UV upturn, in which the flux increases towards shorter
UV wavelengths.

   Model color-magnitude diagrams and corresponding integrated spectra (for 
various values of age, metallicity, helium enrichment, mass loss efficiency, 
initial mass function, and the HB mass dispersion factor) are available on 
S.Y.'s world wide web site http://shemesh.gsfc.nasa.gov/model.html.

\end{abstract}


\section{Introduction}

   The evolutionary population synthesis (EPS) technique, first introduced by 
Tinsley (1968; see \cite{t80} and \cite{bc93} for review), has its fundamental 
basis in stellar evolution theory. 
   Unlike the static population synthesis\footnote{Alternatively, 
``optimized (or optimizing) synthesis'' (\cite{o87}).} 
technique (e.g. \cite{mv68}; \cite{st71}; 
\cite{f72}; \cite{o76}; \cite{p85b}; see \cite{o87} for review) that adopts 
physical constraints from the observational data, EPS relies on theoretical 
constraints, most importantly the evolutionary time scale of stars.

   Some significant advantages of EPS include the following:
(1) the theoretical results can provide physical understanding $when$ the 
results fit the observations reasonably because the solution is unique, 
(2) it allows us to understand not only the
present observable quantities but also the evolution of those quantities,
(3) it is easy to implement the theoretical predictions into the code, and 
(4) it is easier to investigate the sensitivity of the result to input 
assumptions that are used in the modeling.
   While aspects (1) and (2) have been widely recognized, the significance
of aspects (3) and (4) is still poorly understood.

   As an example of aspect (3), EPS enables us to include hypothetical 
stars that are predicted to exist and significant contributors to the 
integrated flux but rarely observed.
   For instance, several stellar evolution theory groups have predicted
that, under certain conditions, low mass core helium-burning stars 
(horizontal branch - HB - stars) become UV-bright before they become white 
dwarf (WD) stars (\cite{dp88}; \cite{gr90}; \cite{ct91}; 
Horch, Demarque, \& Pinsonneault 1992).
   These UV-bright stars, the so-called slow blue phase
(SBP) stars (\cite{hdp92}) which include both AGB-manqu\'{e} stars
(\cite{gr90}) and post-early-AGB (PEAGB, \cite{ct91}) stars, can be 
realistically taken into account only through the EPS technique. 
   Yi, Demarque, \& Kim (1997, hereafter YDK) provided a detailed physical 
description of the UV-bright phase.

   On the other hand, the reliability of EPS is restricted by uncertainty in 
the stellar physics. 
   The SBP stars give an example again. 
   They were unknown until recently. 
   Thus, early EPS studies were not able to include them, despite their 
probable significance in the UV study. 
   However, such pure theoretical predictions from stellar astrophysics may 
not be realistic: such stars may not exist. 
   Then, the EPS using such theoretical constraints would result in wrong 
conclusions.
    Therefore, in earlier times when many of the stellar evolutionary 
phenomena other than the main sequence (MS) and red giant branch (RGB) 
were unclear, the static population synthesis technique was more popular.

   However, there has been remarkable progress in the advanced stellar
evolution theory during the past few decades, especially for the core
helium-burning phase (e.g. \cite{r73}; Sweigart, Mengel, \& Demarque 1974; 
\cite{s87}; \cite{ld90}; \cite{ct91}; \cite{hdp92}; 
Dorman, Rood, \& O'Connell 1993; \cite{fag94}; YDK).
   Therefore, many groups are now including advanced phases of stellar 
evolution in EPS 
(e.g. \cite{np85}; \cite{mb93}; Bressan, Chiosi, \& Fagotto 1994; \cite{b95}; 
Dorman, O'Connell, \& Rood 1995; \cite{yado95}; \cite{pl97}).
   Many of them are mainly aiming at solving the ``UV upturn problem'', where 
the UV upturn is defined as the increasing flux in the UV (1,000 -- 2,500 \AA)
with decreasing wavelength, as found in giant elliptical galaxies
(\cite{cw79}; \cite{f83}; \cite{b88}). 
   
   The most serious concern about such elaborate population studies is that
the effects of input parameters on the resulting UV flux are very poorly
understood.
   Unlike population synthesis in the optical band, UV population
synthesis for old stellar systems, such as elliptical galaxies,
is very sensitive to various detailed physical assumptions.
  This is because most candidates for UV sources are highly evolved stars whose
properties are governed by several input parameters that have not yet been
determined very well.
   Among such parameters are mass loss and its distribution, galactic helium 
enrichment (\DYDZ), and initial mass function.
   In particular, as we will show later, the assumptions about mass loss and 
its distribution influence the magnitude of the resulting UV flux 
significantly.
   Many recent population synthesis studies have been presented to the 
readers without addressing the sensitivity of their results
and conclusions to alterations in the adopted input parameters.
 
   The main goals of this paper are to illustrate the sensitivity of the
model UV spectrum to various input parameters, and to investigate the 
plausibility of UV-bright core helium-burning stars as the major UV sources 
in giant elliptical galaxies.
   This paper complements previous works in the literature.
   Greggio \& Renzini (1990) explored the sensitivity of the UV modeling to 
\DYDZ and to the metallicity-dependence of mass loss.
   The effects of age and metallicity have been investigated by several
groups, among which the works of Bressan et al. (1994), Dorman et al. (1995),
and Tantalo et al. (1996) are noteworthy.
   Although these groups and others (most notably the empirical works done
by the HUT team; \cite{fd93}; \cite{bfdd97}) have tested the plausibility of
core helium-burning stars as the main UV sources in giant elliptical
galaxies before, we found it necessary to investigate it within a fuller
parameter space, in those cases where several input parameters' influence are 
still poorly known.

   Realistic synthetic HB, based on gaussian mass dispersion on the HB, 
has been used in our population synthesis.
   Although Park \& Lee (1997) used the same technique in their recent study
for the first time, ours is the first to combine the mass loss as a 
function of age and metallicity and gaussian mass dispersion.
   We will show why realistic HB treatment is important.
   We also would like to demonstrate that intricate astrophysics must be
considered in estimating mass loss and in incorporating the estimates into
the population synthesis code, even for the most simplistic approach
(e.g. Reimers' mass loss formula).
   
   In order to distinguish the effects of these input parameters
from others, all the models in this paper are based on simple model
populations in which all stars are born at the same time with the same chemical
composition.
   {\it This study focuses on the sensitivity of the UV modeling to the adopted
input parameters, and, thus, does not include a detailed comparison between
models and observational data.}
   The following paper (Yi, Demarque, \& Oemler 1997) in the series is mainly 
dedicated to such comparisons.

\section{Stellar Spectral Library Construction}

   An obvious source of uncertainty in EPS is the stellar spectral library 
used to convert the theoretical quantities (such as \Teff, \logg, and
radius $R$) into observable ones (e.g. magnitudes, colors and spectral line 
strengths) (Charlot, Worthey, \& Bressan 1996).
   Galaxy models have been based on either empirical stellar 
spectral libraries which are highly incomplete in the super metal-rich
($Z >$ \Zsun) regime and lack UV data, or on theoretical libraries that do not 
reproduce observed stellar spectra precisely, and have not been fully 
verified empirically at UV wavelengths. 

   There are several comprehensive empirical libraries 
(e.g. \cite{gs83}; \cite{p85a}; \cite{sc92}) in the optical through 
near-infrared (IR) bands, and they have been used in many population 
synthesis studies.
   In general, they are adequate for MS through RGB stars and 
in the solar abundance regime.
   However, elliptical galaxies are believed to be metal-rich, and no 
adequate empirical library exists for super metal-rich ($Z >$ \Zsun) 
stars. In particular, the lack of empirical spectra of evolved, metal-rich
stars, such as SBP (slow blue phase) stars, is a serious drawback for UV 
studies.
   The best UV spectral library is derived from IUE observations 
(Fanelli, O'Connell, \& Thuan 1987; \cite{fan92}); however, it lacks super 
metal-rich ($>$ \Zsun) samples, and does not cover the far UV near the 
$Lyman$ limit that is crucial to the UV upturn study.
 
   On the other hand, theoretical spectral libraries suffer from insufficient 
atomic data and physical understanding.
   Even the most recent models may be inadequate,
especially for cool stars (\cite{bk92}; \cite{m93}; \cite{k93}; \cite{gj94}).
   These theoretical libraries have to be validated by acquiring as
much empirical data as possible. But, the validation in the UV wavelength
is possible only using the space facilities which have limited and competitive
observing time.

   While both empirical and theoretical libraries have problems, a great 
advantage of using the theoretical spectral library is that we can clearly 
see the effects of various parameters, such as temperature, metallicity,
and surface gravity, in the models. Moreover, the recent Kurucz library covers
the high metallicities and high temperatures that are crucial to highly evolved
stars and thus to UV studies, but that are not available empirically. 
   Thus, a theoretical library has been chosen, assuming that the 
uncertainty in the model spectra is still
smaller than the effects of temperature and metallicity that cannot be 
studied adequately using empirical libraries.

   The foundation of the spectral library in this study is the Kurucz library 
(1992). 
   It covers \feh = $-$4 -- $+$1, \Teff = 3,500 -- 50,000 K,
and \logg = 0.0 -- 5.0 (see Kurucz (1992) for details of the grids). 
   Additional spectra have been constructed for
the stars with high \Teff and/or high \logg using the 1995 version of the 
Hubeny spectral synthesis code, TLUSTY178 (see \cite{h88} for details). 
   The effects of \Teff and metallicity are demonstrated in Figures 1 -- 2.
 
   Since the metallicity effects are small when \Teff $\gtrsim$ 50,000 K, 
all models for \Teff $>$ 50,000 K have been constructed for $Z$ = \Zsun. 
   The stellar spectrum for each star is constructed using linear 
interpolation in metallicity and surface gravity. 
   However, for \Teff, the population synthesis code
finds the model spectrum with the closest temperature in the library instead
of interpolating in the temperature grid in order to save CPU time. 
   For example, the code will 
use model spectra of \Teff = 3,500 K and 3,750 K for the stars of \Teff = 
3,400 K and 3,650 K. This approach slightly overestimates the contribution 
from such cool stars in the composite spectrum in the IR because
the minimum \Teff in the Kurucz library is 3,500 K. However, this effect
is negligible in the UV and optical range because only a small number of
stars are that cool and such cool stars contribute little to the integrated
light in the UV to optical range.

\section{Construction of synthetic CMDs and SEDs}

   EPS consists of two steps, constructing model 
color-magnitude diagrams (CMDs) and spectral energy distributions (SEDs). 
   It will be shown later in this study that a careful synthetic CMD
construction including a realistic HB is crucial, 
especially for UV population synthesis. 
   The model CMD provides basic information, such as \feh, \Teff, \logg, 
and the radius of the star $R$. 
   Since a model stellar spectrum is defined only by \feh, \Teff, \logg, 
and $R$, i.e.
\begin{equation}
f_{\lambda}\, =\, f([Fe/H],\,T_{eff},\,\log\,g,\, R),
\end{equation}
the convolution of this information with the spectral library generates a
model composite spectrum.

   The core helium-burning phase in the population synthesis requires a 
complicated treatment involving a synthetic HB construction. Therefore the 
core hydrogen-burning phase (MS and RGB) and the core helium-burning phase 
- strictly speaking, post-RGB: HB, asymptotic giant branch (AGB), slow
blue phase (SBP), post-AGB (PAGB), and white dwarf (WD) - have been dealt 
with separately.
   Then, the total flux \Flam is the sum of the flux from the core 
hydrogen-burning stars $F_{\lambda}^{H}$ and that from the 
core helium-burning stars $F_{\lambda}^{He}$ 

\begin{equation}
F_{\lambda}\, =\, F_{\lambda}^{H}\,+\,F_{\lambda}^{He} .
\end{equation}

   The foundations of the model CMD construction are the Yale Isochrones 1996 
(\cite{d96}) and Yi et al.'s post-RGB tracks (YDK).

\subsection{Core hydrogen-burning phase: MS and RGB}

   The core hydrogen-burning phase is relatively easier to deal with because
both MS and RGB are well understood. 
   The Yale Isochrones 1996 provides all the necessary information of
the core-hydrogen burning stars for the population synthesis. 

   The initial mass function (IMF) is assumed to follow a conventional power 
law as follows,
\begin{equation}
\frac{d\,n_{i}}{d\,M_{i}} = \alpha\,M_{i}^{-(x+1)}
\end{equation}
where $M_{i}$ is the stellar mass, $n_{i}$ is the number of stars in the 
$i$th mass bin, $\alpha$ is a normalization constant, and $x$ is 
the IMF slope. 
   Three values of $x$, namely -1, 1.35, and 3, are chosen to see the effects 
of IMF on the model CMDs and SEDs; the models with $x = 1.35$ 
(\cite{s55}; \cite{l92}; \cite{l95}) are considered to be standard.

   The flux from all the core hydrogen-burning stars, on the MS and RGB,
$F_{\lambda}^{H}$, is
\begin{equation}
F_{\lambda}^{H} = \sum_{i}\,n_{i}\,\,f_{\lambda,i}
\end{equation}
where
\begin{equation}
n_{i} = \alpha\,\int_{M_{i}-\delta M}^{M_{i}+\delta M}\,M^{-(x+1)}\,dM .
\end{equation}

   The normalization constant $\alpha$ has been set to make the mass of the
model galaxy $10^{12}$~\Msun. This large number avoids any stochastic
effect from rare bright stars. One of the nice features of the population
synthesis of giant elliptical galaxies is that giant elliptical galaxies
are composed of a sufficiently large number of stars that we can avoid
a serious stochastic effect where statistical fluctuations of a few bright 
stars can influence the SED. 
   On the contrary, population synthesis for small galaxies and star
clusters suffers from the stochastic effect.
   Therefore, such studies must be approached with caution.

\subsection{Mass loss estimation}

   Mass loss is one of the most important, but least understood processes 
in stellar astrophysics. 
   It has a dominant influence on the UV evolution of galaxies 
because the mass loss determines the mean \Teff of core helium-burning stars 
that are the major UV sources in old stellar systems, as will be shown later.

   Reimers' empirical formula of mass loss
(\cite{r75}) 
\begin{equation}
\frac{d M}{d t} = -4\,\times\,10^{-13}\,\eta\,\frac{L}{g\,R},
\end{equation}
where $L$ is the luminosity, $g$ is the surface gravity, and $R$ is the radius,
has been used in this study.
   The mass loss efficiency parameter, $\eta$, 
an empirical fitting factor, was originally chosen to be unity based on the 
study of 16 metal-rich (Population I) red giant stars (\cite{r75}). 
   However, $\eta$ has been reported to vary remarkably from 0.25 to 2 -- 3 
(\cite{d86}; \cite{kr78}; \cite{r81}).
   While the estimated $\eta$ for metal-rich stars varies significantly,
the estimated range of $\eta$ for metal-poor stars seems to be much
better determined. Most studies on metal-poor stars suggest
$\eta$ = 0.3 -- 0.6 (\cite{am82}; \cite{ma82}; \cite{r81}; 
Lee, Demarque, \& Zinn 1994).

   One of the best ways to estimate the mass loss in metal-poor stars uses
the fact that the mass of the RR Lyrae variables (a type of HB star)
and the mass of the red giants can be determined independently. 
   In principle, the mass loss on the RGB can be deduced from the mass 
difference between red giants and HB stars.
   However, a precise, empirical mass estimation of ordinary HB stars is 
difficult to achieve.
   Several pulsation studies showed that the mass of a particular type
of RR Lyrae stars (RR-d double mode variables) can be reasonably estimated
(\cite{p73}).
   The mass of RR Lyrae stars can also be measured via stellar evolution 
theory by fitting the observed HB morphology with synthetic HBs 
(Yi, Lee, \& Demarque 1993). 
   For a long time, there existed an inconsistency between the evolutionary 
mass of RR Lyrae variables and their pulsation mass, in the sense that the 
pulsation masses were lower (Petersen 1973; Lee, Demarque \& Zinn 1990).  
   This discrepancy raised doubts in the minds of many variable star 
researchers about the validity of the HB evolutionary models
(e.g. Clement, Kinman, \& Suntzeff 1991).  
   Accordingly, the true mass loss on the RGB was unclear. 
   However, the mass estimates from the two independent methods are now in 
good agreement when improved physics including the OPAL opacities is used 
(Kovacs, Buchler, \& Marom 1991; \cite{c91}; \cite{yld93})\footnote{
There still may be inconsistency in the estimated masses of HB stars.
Recent spectral analyses (e.g. Moehler, Heber, \& de Boer 1995; 
de Boer, Tucholke, \& Schmidt 1997) of hot stars on the blue tail of M\,15
report that the mass estimation is substantially lower than what canonical 
HB models suggest, raising a question either about the origin of the hot 
stars on the blue tail or the validity of stellar evolutionary calculations. 
On the other hand, de Boer et al. note that ``the likely cause for the
discrepancy of the mass values lies in the low gravities derived from
Balmer line profiles.'' (see also \cite{cac95})}.
   Because of this agreement, we can reasonably estimate the amount of mass 
loss experienced by the red giant progenitors of double mode RR Lyrae stars,
once the masses of red giants are provided.
   The masses of the red giants as a function of age and chemical composition 
can be reliably deduced from the stellar evolution theory.

   In order to redetermine $\eta$ for metal-poor stars in this study, we
adopted the evolutionary mass of RR Lyrae stars in metal-poor globular 
clusters, M15 ($Z$ = 0.0001) and M3 ($Z$ = 0.0004), from Yi et al. (1993) 
and the mass of red giants from the Yale Isochrones 1996. 
   It should be emphasized that $\eta$ estimation based on 
the HB morphology, that is, on the mass estimates of RR Lyrae stars,
must use the new mass estimates of RR Lyrae stars after the RR Lyrae
mass discrepancy has been removed. If the old, lower mass is used, 
the $\eta$ would be overestimated by almost a factor of two.
 
   The mass loss (Table 1 -- 4) has been calculated using Reimers' formula 
which is applied to MS -- RGB stellar evolutionary tracks that have been 
constructed without mass loss. 
   Several evolutionary tracks of different masses  were selected for the
mass loss estimation in order to approximately cover the range of 1 -- 25~Gyr 
of age.
   For example, in the case of $Z$ = 0.02 ($\approx$ \Zsun), the tracks 
of 1.8~\Msun (lifetime on the MS -- RGB $\approx$ 1.3~Gyr), 1.4~\Msun 
(3.3~Gyr), 1.0~\Msun (11.1~Gyr), and 0.8~\Msun (31.5~Gyr) were used to
estimate the mass loss at 1.3, 3.3, 11.1, and 31.5~Gyr, and linear 
interpolation was carried out within this age range.
   Therefore, in a 10 billion years old stellar system with ($Z$, $Y$) =
(0.02, 0.27) where most stars on the MS through RGB are slightly heavier than 
$1.0 $\Msun, Reimers' formula suggests that stars experience mass loss of 
about 0.119 (if $\eta = 0.3$) -- 0.396~\Msun (if $\eta = 1.0$) (see the 
second top left panel in Figure 4 and Table 1 -- 4) depending on the adopted 
$\eta$.
   We assume the mass loss becomes noticeable when $L \geq$ \Lsun. 
   Also, the mass loss has been forced to stop in the numerical calculations
either when the star reaches the helium core flash stage or when it causes 
the envelope mass \Menv to be smaller than 0.01~\Msun (as the star loses mass 
and as the core mass \Mcore grows).
   
   It has been suggested that  \Menv cannot be infinitesimally small
(\cite{smd74}) for the following reason. A certain mass is required to be
present in the stellar envelope in order to exert sufficient gravitational
pressure on the degenerate core of an RGB star and to cause a helium core
flash (we call this $the$ $minimum$ \Menv $hypothesis$). 
   This is how HB stars are born.
   It is believed that nearly all the energy produced by the helium core 
flash is consumed in removing the degeneracy of the core and that there is not 
enough energy to expel any significant amount of envelope material into space 
(\cite{t70}; \cite{cd80}; Cole, Demarque, \& Deupree 1985).
   The minimum \Menv has been suggested to be approximately 0.01 -- 0.02~\Msun
(Hayashi, Hoshi, \& Sugimoto 1962; \cite{rw70}; \cite{smd74}).
  
   However, it is still debatable whether such a minimum \Menv exists 
considering that there are hot (low-mass) field subdwarf stars (see 
\cite{smd74} for review) and low-mass UV-bright stars in NGC\,6791 
(\cite{ku92};  Liebert, Saffer, \& Green 1994; \cite{kr95}) that seem to have 
an envelope of \Menv $\lesssim$ 0.01~\Msun. 
   In particular, D'Cruz et al. (1996) recently suggested that such a 
minimum \Menv does not exist. Instead, they claimed that red giants with a 
very smaller envelope mass experience helium-core flash after leaving the tip 
of the RGB quietly.
   We do not include such models in this paper, but we believe that their 
finding is quite plausible and deserves further investigations both to verify 
it with thorough theoretical calculations and to explore the effect of it to 
the population synthesis.
   While the value of the minimum \Menv is not yet clear,
0.01~\Msun has been chosen in this study in order to approximately 
accommodate the hot stars in NGC\,6791 and the hot field subdwarf stars.

   Much of the mass loss is believed to occur during the bright phase of the 
RGB because it is sensitive to the stellar radius $R$, and $R$ increases 
dramatically as a red giant evolves along the RGB. 
   If the minimum \Menv hypothesis is removed, a larger mass loss would be
allowed for some stars and the synthetic HB would contain hotter stars.
   However, the absence of such hot HB stars in Galactic globular clusters
in general supports the minimum \Menv hypothesis\footnote{
While there are several clusters that are known to contain quite hot HB stars
(e.g. M\,15, NGC\,6752, NGC\,1904, \& $\omega$\,Cen), only a few of such hot 
stars are believed to have an \Menv smaller than 0.01~\Msun.
   If such hot stars with \Menv $<$ 0.01~\Msun are remnants of the single 
stellar evolution, they may provide counterexamples to the minimum \Menv 
hypothesis. 
   However, very low mass envelope HB stars might instead be created by mass 
transfer in binary systems (Mengel, Norris \& Gross 1976), or by close 
interactions in dense parts of globular clusters (\cite{bai95}; \cite{fusi93}).
  In this paper we have considered only the evolution of single, 
non-interacting stars.}.
   Table 5 lists the estimated mass of the RR Lyrae variables $M_{RR}$, 
the mass of the RGB stars at the tip $M_{RG}$, and the best fitting $\eta$
for the metal-poor stars. 

   The estimated $\eta$ depends on the assumed ages of the globular clusters 
used in this study. 
   We have chosen 12 -- 16~Gyr as an acceptable age range of the Galactic 
globular clusters M15 and M3 following Chaboyer, Demarque, 
\& Sarajedini (1996). 
   The line with asterisks in the top 
panels of Figure 3 is the locus of the mass loss in the adopted age range, 
and this yields the estimated mass loss efficiency of 
$\eta \approx$ 0.2 -- 0.5, which is in good agreement with the estimates
from the previous studies of metal-poor stars as introduced above.
   Previous studies (see above) seem to suggest that estimated $\eta$'s for 
metal-poor stars are more consistent with one another than for metal-rich 
stars, in which $\eta$ varies from 0.25 to 2 -- 3.
  Despite the large uncertainty, this seems to suggest a metallicity-dependence
of $\eta$ in the sense of increasing $\eta$ with increasing metallicity
if, as suggested by Reimers (1975), $\eta \approx$ 1 for $Z$ = \Zsun.

   Figure 4 shows the mass loss estimates, \DM, for metal-rich stars. 
   There are two general trends to be noted in Figures 3 -- 4. 
   Firstly, \DM generally increases as age increases mainly because 
\MRG decreases with increasing age and the lifetime of a red giant  
($t_{H}$) becomes significantly longer as its mass (\MRG) decreases. 
   Secondly, \DM reaches a maximum when \MRG becomes fairly low. 
   This is because, below this mass, the stellar evolutionary pace is
slower than the stellar core growing pace.
   The mass loss is a product of the two competing time scales, the 
evolutionary time scale on the RGB and the core growth time scale. 
   Figure 5 shows that \DM increases as \MRG decreases until a certain mass
(\MRG $\approx$ 1~\Msun in case of $Z$ = \Zsun) is reached because $t_{H}$ 
increases as \MRG decreases and \DM obviously increases as $t_{H}$ increases. 
   Below this mass, however, the evolutionary pace is much slower than the 
core growing pace, so before the star reaches the tip of the RGB, its core
becomes large enough to cause \Menv $\leq 0.01$~\Msun. 
   Thus a further evolution on the RGB is halted and 
the post-RGB phase begins.
   The post-RGB stars with Menv $<$ 0.01~\Msun are assumed not to become
HB stars but to become WD stars following the $M = 0.453$~\Msun track of
Sweigart et al. (1974).   
   As discussed above, it is possible that some red giant stars with total 
masses very nearly equal to the helium ignition core mass will start core 
helium burning while or after crossing from red to blue giant in the 
HR-diagram (\cite{dm71}; \cite{cc93}; \cite{ddro96}).  
   In this crucial mass range, the mass loss estimation is highly sensitive 
to how the mass loss takes place (which is poorly known), and thus it has to 
be carried out with much caution.
   The same phenomenon exists even if the minimum \Menv hypothesis is removed.

   One can avoid such complications by including mass loss in the stellar
evolution calculations from the beginning, if the mass loss mechanism were
better understood. 
   However, the current standard models do not yet take into account mass 
loss simply because the understanding about mass loss is still primitive.
   Reimers' formula suggests most mass loss takes place near the end of 
the RGB where surface gravity becomes smallest. 
   Moreover, theoretical hydrodynamical models (\cite{bw91}; 
Willson, Bowen, \& Struck 1996) 
suggest that mass loss on the RGB is an even more abrupt phenomenon near the 
end of the RGB than predicted by Reimers' formula. 
   Willson (1996, private communication) even suggested that it would be 
closer to truth to remove a certain mass from the star at the tip of the RGB 
than to include Reimers' formula in the stellar evolutionary calculations. 
   If this is true, whether a mass loss formula is included in the 
construction of the tracks or not causes little difference on the stellar 
evolution and on the estimated mass loss.
   It may even be prudent not to include an arbitrary mass loss formula 
in the evolutionary calculations untill a more reliable mass loss formula 
becomes available.

   The theoretical hydrodynamical models suggest that mass loss takes place 
when the star reaches a certain critical luminosity (the ``cliff\,''), which 
is a function of mass and metallicity. 
   Below the cliff, the mass loss rate is too low to be observable; and 
above the cliff mass loss takes place so rapidly that the star is unlikely 
to be observed in this phase of evolution. 
   It is significant that their models precisely reproduce 
Reimers's formula with $\eta$ = 1 for $Z$ = \Zsun.
   Moreover, they explain the positive metallicity-dependence 
of $\eta$ which observations may be indicating.
   While their models have not yet been parameterized for population synthesis,
Willson et al. (1996, private communication) advise using Reimers' 
formula with variable $\eta$ in the sense that $\eta$ increases as metallicity
increases.
   We call this {\it the variable $\eta$ hypothesis}. 
   Table 6 shows the suggested $\eta$ for different metallicity.

   The population synthesis models in this study have all been constructed
using values of $\eta$ = 0.3, 0.5, 0.7, and 1.0 that roughly span the range of 
empirical and hydrodynamical estimates. 
   This will allow us to investigate the effects of 
mass loss efficiency on the model CMDs and SEDs. 
   The impact of the variable $\eta$ hypothesis, which has been investigated 
earlier (\cite{gr90}) for single abundance models, will be revisited for 
the composite galaxy models in our next paper (Yi, Demarque, \& Oemler 1997).

\subsection{Core helium-burning phase: post-RGB}

   We adopt the evolutionary tracks of core helium-burning phase mainly 
from YDK. Since YDK do not provide post-AGB (PAGB) tracks, the PAGB tracks of
Kiel group (\cite{s79}; \cite{s83}; \cite{bs90}) have been adopted.
   The YDK HB tracks in this study are the short version (plus symbols in 
Figures 1 -- 2 of YDK) of the full tracks (as solid lines in their 
Figures 1 -- 2). 
   The time step on each HB track is 5 Myr and the mass step is
approximately 0.04~\Msun, with smaller steps near the blue end of the HB
as listed in Table 1 of YDK.
   The masses of the PAGB stars and WD's and the process that determines
them are yet to be understood. 
   Meanwhile, Kiel group's models of $M = 0.546$, 0.565, and 0.598~\Msun have 
been used in the actual population synthesis. 
   One of these three models was chosen as the progeny of an HB star
(almost arbitrarily) depending on the mass of the HB star; that is, 
0.546, 0.565, and 0.598~\Msun models were chosen for the HB stars with
$M \leq 0.56$, $M = 0.60$, and $M \geq 0.64$.
   However, the choice of the model does not affect the resulting model 
integrated spectrum much because PAGB stars are not likely to be the major 
UV sources when a galaxy shows a strong UV upturn, as will be shown in 
Section 4.1.
   
   YDK (Figures 1 -- 2 of YDK) show that late stellar evolution in the core 
helium-burning phase, especially in metal-rich galaxies, is very complicated. 
   And, such evolved, metal-rich core helium-burning stars are
potentially significant UV sources.
   Therefore, in order to fully incorporate the complete evolution of 
core helium-burning stars in the population synthesis code, a physically
plausible treatment of the phase is crucial.

   The modified gaussian mass distribution (\cite{ldz90}), 
first invented by Rood (1973) and later modified and elaborated on by 
Lee et al., has been assumed in this study for the
core helium-burning phase as follows:
\begin{equation}
\Psi(M)\, = \,\Psi_{0}\, [\, M\, - \,(\MMHBeq\,-\,\Delta M)\,]\,
(M_{RG}\,-\,M) \,\, exp\,[\,-\,\frac{(\MMHBeq\,-\,M)^{2}}{\sigma^{2}}]
\end{equation}
where $\sigma$ ($\equiv 2 \sigma_{sd}$) is a mass dispersion factor, 
$\Psi_{0}$ is a normalization factor, and \MMHB ($\equiv$ \MRG $-$ \DM) is 
the mean mass of HB stars. 
  Three values of $\sigma$ (0.01, 0.06, and 0.99 in~\Msun) have been chosen to 
represent distributions similar to a delta-function, a 
globular-cluster-function (\cite{ldz90}), and a near-constant-function, 
respectively. 
   While the true value of $\sigma$ is still quite uncertain (\cite{r73};
\cite{sn83}; \cite{ldz90}; \cite{w92}; \cite{c93}), we have chosen 
$\sigma$ = 0.06~\Msun as the ``standard'' value in this study following 
empirical (but, with no physical basis) choices of Rood (1973) and 
Lee et al. (1990).
   The shape of the functions are shown in Figure 6.
   In actual calculations, we first estimate the mass loss as a function of
age and metallicity, then, apply the gaussian distribution on the 
mass loss\footnote{
   Recently, J$\mbox{\o}$rgensen \& Thejll (1993a) and D'Cruz et al. (1996)
attempted to explain the width of the HB of Galactic globular clusters
by means of a spread in $\eta$ (mass loss efficiency) rather than a spread
in mass loss.
   J$\mbox{\o}$rgensen \& Thejll (1993a) noted that ``star-to-star variations
of $\eta$, between $\eta$ = 0.0 and $\eta$ = 0.7 can explain the HB
morphology in typical globular clusters''.
   It would be very interesting to investigate the effect of such HB treatments
to the UV population synthesis.}.
   Assumption of a (modified) gaussian function for the HB mass distribution
has been questioned (\cite{sn83}; \cite{r90}; \cite{fer92}; \cite{dddf96}). 
   However, the single-gaussian distribution has been very successful in 
reproducing the observed HB morphology in Galactic globular clusters 
(\cite{ldz90}; \cite{ldz94}; \cite{dddf96}; \cite{c93}).

The total number of core helium-burning stars, $N_{He}$, is as follows:
\begin{equation}
N_{He} \,=\, \overline{t_{He}}\, R_{He} 
\end{equation}
where $\overline{t_{He}}$ is the lifetime of the core helium-burning star of 
$M$ = \MMHB, and $R_{He}$ is the rate at which stars leave the
RGB for the core helium-burning stage. 
   So, at a fixed time, there will be $N_{He}$ core helium-burning stars on 
all post-RGB phases.
   The rate, $R_{He}$ in number of stars per unit time interval $t_{1} - t_{2}$
is defined to be
\begin{equation}
R_{He}\,=\,\frac{\alpha}{t_{1}\,-\,t_{2}}\,\,\int_{M_{GB}(t_{2})}^{M_{GB}(t_{1})}\,M^{-(x+1)}\, dM.
\end{equation}

   The calculated probability, $\Psi(M)$, is applied to each evolutionary 
track according to its mass $M_{i}$. 
   There is a different list of stars for different chemical compositions, 
and each track has a number of evolutionary points with evolutionary time 
scale information\footnote{Evolutionary time scale: how long a star stays 
within a box of given size in the CMD.}. 
   Thus, the number of stars at the $j$th evolutionary point on the $i$th 
track will be
\begin{equation}
n _{i,j} \,= \,\Psi(M_{i})\, N_{He}\, \frac{t_{i,j}}{t_{i}}
\end{equation}
and
\begin{equation}
t _{i} \,= \,\sum_{j} \, t_{i,j}
\end{equation}
where $\Psi(M_{i})$ is the probability of the star of $M = M_{i}$ 
to be present, $t_{i,j}$ is the time spent by the $i$th star 
in the table near the $j$th position in the CMD and $t_{i}$ is the lifetime 
of the $i$th star in the post-RGB phase.
   The sum of the probability $\Psi(M_{i})$ should be unity:
\begin{equation}
\Psi\, =\, \sum_{i}\, \Psi(M_{i})\, =\, 1 .
\end{equation}

Then the total flux from the core helium-burning stars, $F_{\lambda}^{He}$, is
\begin{equation}
F_{\lambda}^{He}\, =\, \sum_{i}\!\sum_{j}\, f_{\lambda,i,j}\,\, n_{i,j} ,
\end{equation}
where $f_{\lambda,i,j}$ is the flux from the $i$th star in the list of
post-RGB stars near the $j$th evolutionary point.

\section{Effects of input parameters}

   Spectral evolution of galaxies in the UV is much more uncertain than
that in the optical band because it is more sensitive to 
the evolution of post-RGB stars which are less well understood than 
the MS -- RGB stars that dominate the optical flux.
   It is important to understand first the effects of input parameters on the
resulting CMD and UV SED in order to answer the questions related to
the spectral evolution of the galaxy, including the UV upturn phenomenon.

   Primary parameters affecting the UV flux are age and metallicity. 
   These are the most fundamental quantities in describing stellar populations,
because they predominantly determine the \Teff's of post-RGB stars.   
   We would like to investigate whether metal-rich old stellar systems, such as
giant elliptical galaxies, are capable of exhibiting a pronounced UV flux.
   Secondary parameters include mass loss efficiency, galactic helium 
enrichment parameter, form of the dispersion in mass loss, and the slope
of IMF, while these parameters are not necessarily less important in
producing UV flux than the primary parameters.   
   We assume  in this study that all stars in a model stellar system form 
in an instantaneous starburst. 
   
\subsection{Age and metallicity: primary parameters}
   
   The effects of age and metallicity in old stellar systems have recently 
been widely recognized through studies of Greggio \& Renzini (1990), 
Bressan et al. (1994), and of Dorman et al. (1995).
   In this section, we would like to provide a simple explanation of the basic 
conclusions of such previous studies by means of stellar evolution theory.
   We also illustrate that core helium-burning stars are a more
likely major UV source in old stellar systems than PAGB stars.
   Both a higher age and a higher metallicity cause a galaxy to look 
redder in the optical band mainly because MS and RGB stars become cooler as 
they age and as metallicity increases, as is clearly shown in Figures 7 -- 12.
   The effects of age and metallicity on the UV flux are more complicated,
particularly because the major UV sources vary with time. 
   Before we discuss the details, it is important to remember that
we are investigating the effects of age and metallicity while other
parameters, such as $\eta$, are kept constant.
   If other parameters are age-dependent and/or metallicity-dependent
the true effects of age and metallicity become much more complicated.

\subsubsection{Age}

   The major UV sources vary with time. 
   As seen in Figures 7 -- 12, PAGB stars are the only far-UV sources in the 
early ($\lesssim$ 5 -- 10~Gyr, depending on the metallicity) history of galaxy 
evolution, when the majority of core helium-burning stars are still too cool 
to produce any UV light. 
   However, once the galaxy is old enough to develop a substantial number of 
hot core helium-burning stars, those stars dominate the UV spectrum.
   
   In order to investigate the light contribution from various sources,
we first arbitrarily divide the core helium-burning stage into two:
central helium-burning stage and shell helium-burning stage\footnote{The terms,
``core helium-burning'' and ``central helium-burning'', may be confusing.
Readers should pay attention to the different definitions of the terms.}.
   The central helium-burning stage is defined as the stage 
where most of the helium-burning takes place near the center of the star.
   This stage lasts from the zero age HB (ZAHB) to the point where the 
central helium is almost exhausted ($Y_{c} = 0.01$). 
   Similarly, the shell helium-burning stage is defined as the stage that 
stretches from this point until the star reaches either the tip of the AGB or 
the beginning of the WD phase, whichever comes first.
   So the central helium-burning stage is generally referred to as the HB, and
the shell helium-burning stage as the post-HB phase including the evolved HB, 
the AGB and the slow blue phase (SBP).
   These two stages are denoted as crosses and squares in Figure 13.

   Assuming that all the stars in elliptical galaxies have formed in an 
instantaneous burst, Figure 14 shows that central and shell helium-burning 
stars are far more efficient UV sources than PAGB stars in old galaxies.
   The upper and lower panels of Figure 14 show the light contribution
from various evolutionary stages in young and old $metal$-$rich$ model 
galaxies.
   It shows that MS and RGB stars are the dominant sources in the
visible range at all times. 
   The far-UV spectrum is, however, mainly dominated by highly evolved stars. 
   At small ages, PAGB and MS Turn-Off stars are the dominant far-UV and 
near-UV sources (the upper panel), respectively.
   However, such galaxies do not exhibit any significant amount of UV flux, 
as shown in the 5~Gyr old models in Figures 10 -- 12.
   The critical moment which divides the periods in which PAGB stars or
core helium-burning stars dominate the far-UV flux is a  sensitive function
of chemical composition and other parameters (such as $\eta$).
   In the case of the extreme-UV ($\lambda \lesssim 300 $\AA), PAGB stars 
always dominate because of their high \Teff's.
   However, once galaxies become old enough to contain hot HB stars,
shell helium-burning stars dominate the flux in the far-UV 
($\lambda \approx$ 300 -- 1,500 \AA), while both central and shell 
helium-burning stars are equally important in the near-UV ($\lambda \approx$ 
1,500 -- 3,000 \AA).

   The effect of age on the UV flux is relatively simple.
   Figures 15 -- 16 show the UV-to-$V$ colors as a function of age and 
metallicity.
   The model magnitudes are defined as 
$m_\lambda$~$=$~$-2.5$~log~$<\!f_\lambda\!>$ where $<\!f_\lambda\!>$ 
is the mean flux in the bandpass. 
   The $<\!f(1500)\!>$, $<\!f(2500)\!>$, and $<\!f(V)\!>$ are defined by 
averaging the flux within the ranges 1250 -- 1850 \AA, 2200 -- 2800 \AA 
(\cite{dor95}), and 5055 -- 5945 \AA (\cite{a76}), respectively.
   The observational data are from Table 1 -- 2 of Dorman et al. (1995).
   From their table, we selected the clusters for which both m(1500)-$V$
and m(2500)-$V$ are available.
   They are mostly $ANS$ or $OAO$-2 data.
   The UIT data of NGC\,1904 (M\,79) and NGC\,5139 ($\omega$\,Cen) have also
been adapted.
   As a result, three clusters (NGC\,1904, NGC\,5139, and NGC\,5272) have 
two data points in our adapted cluster sample.
   We excluded the IUE data because the IUE aperture is so small that IUE
data do not represent the integrated light properly (\cite{dor95}).
   In case of metal-poor systems (Figure 15), UV-to-$V$ colors become redder 
with age for a while as the MS Turn-Off stars, the major UV sources in young
metal-poor systems, become redder with time, and then they turn around to
become bluer as HB stars get hotter.
   Most metal-poor globular clusters are reasonably matched by the models with
$\eta = 0.5$ but metal-rich globular clusters like NGC\,6388 and 47 Tuc are 
not matched by the same models\footnote{The m(1500)-$V$ of 47 Tuc obtained 
by $OAO$-2 is likely to be the lower (bluest) limit because there is one 
bright PAGB star in $OAO$-2's large aperture (\cite{dor95}).}. 
   This issue of $\eta$ will be discussed in the next section.
   Similarly, metal-rich elliptical galaxies are reasonably reproduced by
the models with $\eta = 0.7$ in Figure 16.

\subsubsection{Metallicity}

   The contribution from each evolutionary stage to the UV flux also varies 
with metallicity because the post-HB evolutionary pattern differs according to
metallicity, as shown in Figures 1 -- 2 of YDK. 
   Since the UV-bright SBP phenomenon becomes far more prominent in more 
metal-rich stars, the contribution from the shell helium-burning stars in 
the far-UV is more conspicuous in metal-rich models (Figure 17). 

   It should be noted that the light contribution from evolved core
helium-burning stars to the total $U$ band flux ($\lambda \approx$ 3,000 - 
4,000 \AA) is not negligible.
   The lower panels of Figures 14 and 17 suggest that substantial amount 
of the total light comes from such evolved stars.
   However, note that the precise quantity highly depends on the adopted 
input parameters.
   Earlier population synthesis studies (e.g. \cite{t72}; \cite{o76}) noticed
that population synthesis models, that did not include such evolved stars
in an adequate way, were predicting much redder $U-B$ colors than observed.
   They suggested that this mismatch could originate from the presence
of young stars in giant elliptical galaxies.
   However, Gunn, Stryker, \& Tinsley (1981) recognized the lack of empirical 
support for the young star interpretation, and suggested blue stragglers and 
what we now call PAGB stars as possible candidates for the observed UV flux.  
   We understand now, as clearly shown in Figures 14 and 17, that a proper
treatment of evolved stars, especially with a realistic mass dispersion
on the HB, enhances the UV -- $U$ band flux significantly, as pointed out
by Burstein et al. (1988) earlier\footnote{The referee pointed out that
the ``excess'' near-UV (U band) light in elliptical galaxies could also be 
due to intermediate temperature (younger) main sequence stars.}.
   This short wavelength region is extremely sensitive not only to age and
metallicity of the galaxy but also to various input parameters, including
mass loss efficiency and the HB mass distribution, as will be shown in the 
following sections.

   The overall effect of metallicity is more complicated than that of age 
because it is a product of several competing effects. 
   Even when we consider only the period in which the core helium-burning 
stars are the main UV sources, it is better to see the metallicity effect in 
two different metallicity domains, (1) $Z <$ \Zsun and (2) $Z \gtrsim$ \Zsun, 

  When $Z <$ \Zsun, UV-to-$V$ flux ratios decrease with increasing metallicity.
   This is a result of the following competing effects.
   (1) The temperature of the HB stars decreases dramatically as 
metallicity increases, (2) more metal-rich stars stay on the MS -- RGB longer
(\cite{gr90}; \cite{jt93b}), but
(3) a larger mass loss of more metal-rich stars for a fixed $\eta$ 
(Figures 3 -- 4) compensates for these two effects (\cite{jt93b}). 
   The phenomenon (2) is mainly caused by the delay of the evolution due
to the opacity effect. 
   Consequently, the phenomenon (2) makes a more metal-rich cluster have more 
massive red giants at a fixed age, but (3) compensates for this effect and 
makes the \MMHB of more metal-rich clusters smaller than that of metal-poor 
clsuters in this metallicity domain. 
   However, the phenomenon in (1) again negates the result of the competition
between that of (2) and (3). 
   The net effect of all these competitions is a decrease in the UV-to-$V$
flux ratios with increasing metallicity which is consistent with observations,
as illustrated in Figure 15.
   In Figure 15, three metal-poor models covering 1 -- 25~Gyr of age are 
compared with globular cluster data.
   It is clear that both observations and models indicate that more metal-rich
systems in this metal-poor regime are redder in these UV-to-$V$ colors.
   It should be noted that this trend (the correlation between the UV-to-$V$ 
ratios and metallicity) is based on the assumption that other parameters
are identical.
   In the case of Galactic globular clusters, it has been suggested that
there is a substantial spread in age among them which influences
the HB morphology and thus the flux ratios.
   A large scatter in the flux ratios in Figure 16 may have been caused by such
differences in cluster parameters.

   In the metal-rich regime $Z \gtrsim$ \Zsun, the evolutionary phenomena are
quite different. 
   Firstly, the higher the metallicity, the faster the MS -- RGB evolution, 
because the luminosity from the hydrogen burning is significantly sensitive 
to the mean molecular weight that is determined by the chemical composition.
   This has been noticed previously by a number of workers (e.g. \cite{gr90};
\cite{jt93b}).
   Under the assumption of a positive \DYDZ ($\approx$ 2 -- 3 in this study),
a higher metallicity means a higher helium abundance.
  A higher helium abundance causes a higher mean molecular weight that
results in a faster stellar evolution (see YDK).
   This causes the \MRG of more metal-rich stars at a fixed age to be smaller.
   Secondly, the increase in mass loss as a function of metallicity 
(Figures 3 -- 4) makes \MMHB of more metal-rich HB stars much 
smaller at a fixed age. 
   In addition to this, even many of the metal-rich HB stars that 
are born as extremely cool stars quickly evolve to become UV-bright stars
(namely, SBP stars in Figures 1 -- 2 of YDK). 
   As a result, more metal-rich galaxies are expected to contain more
UV-bright stars at a given age in this metal-rich regime, given other 
parameters fixed. 
   This is illustrated in Figure 16: the more metal-rich, the bluer. 
   Figure 16 seems to suggest that the galaxy data are better fitted by the
$Z = 0.01$ model than more metal-rich models.
   We believe that this is an artifact that is produced by the comparison 
between the galaxy data and single-abundance models.
   Obviously, galaxies are not composed of stars of a single abundance
only.
   The effect of the composite population treatment will be discussed
in detail in our next paper.

   The data and models in Figure 18 show that metal-poor globular clusters in 
general exhibit bluer m(1500)-$V$ and m(2500)-$V$ colors than elliptical 
galaxies (bottom panel of Figure 18).
   This is mainly because of the opacity effect on the HB stars (in the case
of m(1500)-$V$) and on the MS stars (in the case of m(2500)-$V$).
   For a given age, the opacity effect on the stellar evolution causes 
metal-poor blue HB stars to have higher \Teff's than metal-rich counterparts, 
even though the mean mass of metal-rich HB stars, \MMHB, are likely to be 
smaller (note that, given the other parameters fixed, a lower-mass HB star is 
hotter) according to Reimers' formula (see Figures 3 -- 4 and Table 1 --4). 
   Moreover, even if their \Teff's are identical, the opacity effect 
(via the line-blanketing effect in the spectral synthesis) causes more 
metal-rich HB stars to look much redder than metal-poor HB stars. 
   For the same reason, metal-poor systems show bluer m(2500)-$V$ colors, but 
this time because of the opacity effect on $MS$ $stars$ (\cite{dor95}).
   As a result, metal-poor populations exhibit stronger UV-to-$V$ flux ratios 
than metal-rich populations. 
   However, metal-poor populations with such blue UV-to-$V$ colors do not
exhibit a ``UV upturn'' (the continuously increasing flux with decreasing 
wavelength below 2,500 \AA). 
   Instead, their UV spectra, even for the UV-strong globular clusters,
are rather flat below  2,500 \AA.
   Consequently, globular clusters have redder m(1500)-m(2500) colors than
elliptical galaxies, as shown in the top panel of Figure 18.
   So, it should be said that only elliptical galaxies exhibit a 
``UV upturn'' in a strict sense, as models predict in Figure 18.
   This result disagrees with Park \& Lee (1997)'s recent suggestion.
   Details will be discussed in our next paper.

\subsection{Mass loss efficiency}

   Figures 19 -- 20 show that the magnitude of the UV flux is sensitive 
to the mass loss efficiency $\eta$ in Reimers' formula. 
   A larger $\eta$ results in lower-mass, hotter HB stars.
Therefore, a value for mass loss efficiency should be adopted carefully and 
with sufficient justification. 
   The probable range of $\eta$ seems to be 0.3 -- 1.0 (or even 2) with a 
possible positive metallicity-dependence supported both by observations and
hydrodynamical models, as discussed in Section 3.2.
   Such sensitivity of the UV flux to the mass loss efficiency has significant 
implications. 

   If Reimers' formula is an adequate approximation of reality, 
$\eta$ of stars of $Z \gtrsim$ \Zsun should be near or larger than 0.7 
(this value depends on the assumed metallicity of giant elliptical galaxies)
in order for a metal-rich elliptical galaxy to exhibit a UV upturn within 
a Hubble time\footnote{We assume that a Hubble time is slightly larger
than the age of the oldest Galactic globular cluster estimated using the 
same stellar evolutionary calculations as the ones used in this study. 
This study is based on the stellar models that suggest an age of approximately 
15~Gyr for the oldest Galactic globular clusters (\cite{cds96}).}.
   Old, metal-rich galaxies should have an appreciable number of UV-bright 
stars and a noticeable UV upturn as long as $\eta \gtrsim 0.7$ for metal-rich 
($Z \gtrsim$ \Zsun) stars.
   On the other hand, if $\eta \lesssim 0.7$ for metal-poor stars, as both 
observations and theory suggest, metal-poor stars $cannot$
be major UV sources in elliptical galaxies because they do not
produce sufficient far-UV flux (e.g. m(1500)-m(2500) color) 
within a Hubble time, as Figure 20 shows. 
   This problem becomes worse when even shorter wavelength regions
($\lesssim$ 1,500 \AA) are compared, because the UV spectral energy 
distributions of giant elliptical galaxies have much higher far-UV flux than 
near-UV flux, and this cannot be produced by metal-poor models unless 
$\eta$ is unrealistically high.
   Even if larger $\eta$-models (e.g. 1.0) may reproduce such high 
far-UV flux (bottom panel in Figure 20), the predicted overal spectral shape 
in the UV  does not look similar to what is observed.
   Detailed quantitative study will be shortly presented by Yi et al. (1997).

   Conversely, the magnitude of the UV upturn can be used to set a constraint
on the mass loss efficiency of Galactic globular clusters whose age and
metallicity are independently estimated\footnote{In order for this method
to be reliable, far-UV ($\lesssim$ 1,500 \AA) spectra should be available.}. 
   For instance, 
if we compare the far-UV spectrum of M\,79 (Figure 5 in \cite{dddf96}) with 
Figure 20, the model with $\eta \approx$ 0.5 reproduces the empirical
spectrum best, assuming 15~Gyr for its age and ``standard'' values of model 
parameters ($x$ = 1.35, $\sigma$ = 0.06). 

   Figure 21 illustrates the $\eta$-dependence of metal-poor models.
   As discussed earlier in Section 3.2, low-$\eta$ models ($\eta$ = 0.3 -- 0.5)
reasonably match the observations at an age of $\approx$ 15~Gyrs.
   However, the metal-rich clusters (filled squares, e.g. NGC\,6388) cannot 
be matched by the models with consistent ages.
   For example, if $\eta = 0.3$ even for such metal-rich globular clusters,
UV color fitting would suggest NGC\,6388 is almost 25~Gyrs old, which is 
much higher than the average isochrone-age of globular clusters 
($\approx$ 15~Gyr, \cite{cds96}).
   If a larger $\eta$ is adopted for such metal-rich globular clusters instead,
their colors can  be reproduced successfully at a reasonable age.
   For instance, the age estimate for NGC\,6388 would be $\approx$ 19 
Gyr and 15~Gyr if $\eta = 0.5$ and 0.7, respectively.
   If other complexities are ignored at the moment, this suggests 
$\eta \approx$ 0.7 for $Z \approx$ 0.006 (the metallicity of NGC\,6388), which
is higher than the best fitting $\eta$ for metal-poor clusters.
   This metallicity-dependence of $\eta$ is consistent with the variable-$\eta$
hypothesis discussed in Section 3.2.
   Elliptical galaxies are also better matched by high-$\eta$ models
($\eta \gtrsim$ 0.7), as shown in Figure 22, which is again consistent with 
the variable-$\eta$ hypothesis.
   Note that in this diagram we use m(1500)-m(2500) as the y-axis
because m(1500)-m(2500) describes the UV upturn strength of galaxies better.
   Regardless of metallicity, $\eta$ certainly plays a very important role
in model UV flux in the sense that a higher-$\eta$, at least in the rage
$\eta$ = 0.3 -- 1.0, leads to a higher UV flux.

\subsection{\DYDZ}

   The galactic helium enrichment parameter, \DYDZ, is another important
determinant of the UV flux, not only because it affects the pace of evolution 
of MS and RGB stars (\cite{gr90}; \cite{jt93b}), but also because metal-rich 
stars with a higher \DYDZ become UV-bright more readily (\cite{hdp92}; 
\cite{dro93}; YDK). 
   This phenomenon is more significant when $Z \gtrsim$ \Zsun, with an 
implicit assumption of a positive \DYDZ. 
   Therefore, a metal-rich galaxy with a higher value of \DYDZ
produces a higher UV flux, as shown in Figures 23 -- 24.
   Qualitatively, models with \DYDZ = 2 develop a similar magnitude of UV flux
to that of \DYDZ = 3 at approximately 10 -- 20 \% greater ages.
   Despite such a difference, it is difficult to estimate the true \DYDZ from 
UV magnitudes only because the evolutionary paths for different \DYDZ
are very similar as shown in Figure 24, unless their ages are already known.

   If \DYDZ is not positive (a very unrealistic assumption in the light of 
what is known about nucleosynthesis in massive stars, see e.g. \cite{m91}), 
in other words, if helium does not increase with increasing
metallicity, several complicated effects will compete with one another:
(1) The SBP (slow blue phase) phenomenon will not be significant even in 
metal-rich (but, not unrealistically metal-rich) populations (\cite{dro93};
\cite{bcf94}). This reduces the UV flux from such hot SBP stars.
(2) The evolutionary pace of the more metal-rich stars will perhaps become 
slower because the opacity effect 
will overwhelm the hydrogen luminosity-mean molecular weight effect
(See Eqn. 3 in YDK). This is similar to the
metallicity effect on the evolutionary pace of subsolar-abundance stars;
namely, the more metal-rich, the slower the evolutionary pace on the MS and
RGB if $Z \lesssim$ \Zsun.  This causes a cooler HB.
(3) But, the mass loss will still increase with metallicity regardless of
helium abundance. This will cause HB stars to become hotter.
   The final result of this competition is not trivial to guess. 
Unfortunately, the stellar models of \DYDZ = 0 are not yet available 
for numerical experiments.

\subsection{IMF}

   The slope of IMF, $x$, as defined in Eqn. 3, has a very small effect on
the resulting UV flux as shown in Figures 25 -- 27, even if $x$ differs 
from one galaxy to another significantly.
   A model with a smaller $x$ sends more stars from the MS both to the RGB 
and to the HB and results in an increase in flux in all wavelength regions.
   So the relative flux is not much affected. 
   Many observations have supported $x \approx$ 1.35 
(\cite{s55}; \cite{ms79}; \cite{l92}; \cite{l95}), and thus
we have little justification for adopting a significantly different value. 
   The effect of the difference in various forms of IMF for the very 
low-mass stars on the resulting SEDs is negligible because they contribute 
little to the total light of the host galaxy although they could affect the 
calculated mass-to-luminosity ratio significantly.

\subsection{Mass dispersion on the HB}

   A smooth mass dispersion on the HB has not been seriously taken into account
as an input parameter in earlier population synthesis studies\footnote{
   Dorman et al. (1995) treated the HB as a product of two different values 
of mass loss on the RGB (i.e. bimodal distribution in the mass loss: two peaks
- a red HB clump and a blue HB clump - instead of a smooth distribution)
to understand how large a fraction of hot HB components is needed to account 
for the observed UV flux of elliptical galaxies.
  Their blue HB clump is either by $\delta$-function or by constant-function 
within a certain mass range, both of which are arbitrary.
   Although this is a much improved effort compared to previous works in terms
of the HB treatment, it is still ad hoc because such bimodality is supported
(at least so far) only by a few globular clusters, e.g. NGC\,2808 
(Crocker, Rood, \& O'Connell 1988).
   A single-gaussian mass distribution reproduces the HB distribution
adequately (\cite{ldz90}; \cite{ldz94}), as discussed in Section 3.2.},
although it is very important to the resulting SED in the UV. 
   Its effect is not monotonic.
When a galaxy is young, and has mostly red HB stars, allowing a larger
dispersion in mass leads to a larger number of hot HB stars, which
results in a stronger UV flux (Figure 27). 
   On the other hand, when a galaxy is already old enough to have most of its 
HB stars on the hot side, a larger mass dispersion causes a weaker UV flux
(Figure 28). 
   These effects on the colors are illustrated in Figure 29. 
   As is the case of \DYDZ, it is difficult to choose an optimal
$\sigma$ from the UV colors because models with even very different values
of $\sigma$ follow very similar evolutionary paths (Figure 29) unless
the age of the galaxies is known a priori.

   Because of this complex sensitivity of the UV flux to the assumed
HB mass distribution, a realistic HB treatment is essential to the UV
population synthesis. Oversimplified HB treatments, such as a clump star
assumption\footnote{It assumes that all the HB stars have
the same physical properties, luminosities and temperature, ignoring
the mass dispersion on the HB and the advanced evolution beyond the ZAHB.},
a single-mass assumption\footnote{It assumes that all the HB stars have
the same mass and follow the same evolution.} and a flat 
distribution\footnote{This assumption allows a mass dispersion but 
does not consider any concentration of stars at some mass.}, are inadequate 
for UV studies unless there is a sufficient justification.
   
\section{Summary}

   Recently, several groups have been using the evolutionary population
synthesis (EPS) technique in the hope of understanding the stellar content 
and the evolution of giant elliptical galaxies.
   Most of these studies are aimed at solving the so-called UV upturn
mystery.
   Although EPS in the UV is a powerful technique for these purposes, 
its dependence on various input parameters has not been widely known.
   Therefore, both modelers and model users are vulnerable to
misinterpretations of the results from their population synthesis studies.

   The EPS technique has been described in this study in detail. 
   We show that the model UV flux is very sensitive to several input 
parameters as various earlier studies pointed out.
   These input parameters include not only age and metallicity, but also 
mass loss efficiency $\eta$, galactic helium enrichment parameter \DYDZ, 
and mass distribution on the HB. 
   The effects of these parameters are as follows.

1. Once MS turn-off stars become too cool to dominate the near-UV integrated 
flux, the relative strength of UV flux to visible flux always increases 
with age because a larger number of hotter core helium-burning stars 
develop as a galaxy ages. 
   Before this point, UV-to-$V$ colors become redder with increasing age as
MS stars, the major near-UV sources at low ages, become cooler.
This turning point is sensitive to metallicity and other parameters.
It could be approximately 1~Gyr or even smaller if $Z \gtrsim$ 0.06 but could 
be as large as 10~Gyrs if $Z$ = 0.0004.
   A possibility that giant elliptical galaxies 
have a substantial fraction of very young ($<$ 1~Gyr) stars is ignored
in this study because of lack of empirical support (\cite{o92}; \cite{ber93}).

2. In the metal-rich regime ($Z \gtrsim$ \Zsun), 
metallicity has a positive impact on the UV flux once a model galaxy 
is old enough to have a substantial number of hot core helium-burning stars.
This is caused mainly by the following phenomena in stellar evolution: 
under the assumption of a moderate \DYDZ ($\approx$ 2 -- 3), 
(1) more metal-rich stars evolve faster when $Z \gtrsim$ \Zsun, which results 
in a quicker hot HB development, (2) more metal-rich red giants experience
a larger mass loss, and (3) more metal-rich core helium-burning stars become 
UV-bright stars in the slow blue phase (SBP) more easily.
   This result is in agreement with Brown et al. (1997)'s recent spectral
analysis which suggests that ``models with supersolar metal abundance and
helium best reproduce the flux across the entire HUT wavelength range...''.
   Optical spectra of giant elliptical galaxies suggest $Z \gtrsim$ \Zsun.
   In the metal-poor regime, the metallicity effect is reversed, however.
This is mainly because the opacity effect overwhelms the metallicity-mass loss
relation. 
   In addition, in this metal-poor regime, an increase in metallicity causes 
stars on the MS -- RGB to evolve more slowly and causes the HB to look redder.

3. Because the UV-bright phase (the SBP) of core helium-burning stars is 
positively correlated with helium abundance, a model galaxy with a higher 
\DYDZ exhibits a stronger UV upturn. This agrees with the result of Greggio \&
Renzini (1990). 
   If, as we expect, the true \DYDZ is within the range 2 -- 3, then
the degree of the sensitivity seems moderate.
   The effect of \DYDZ and the cause of the effect are similar to those of
metallicity described above. 
   A model with \DYDZ = 2 would develop a UV flux with a similar magnitude
to that of a model with \DYDZ = 3 at slightly later time (approximately  
10 -- 20 \% of their ages).

4. A larger mass loss efficiency, $\eta$, causes a stronger UV upturn
because it causes a hotter HB. 
The probable range of $\eta$ is 0.3 -- 1 (or higher) with a large uncertainty 
and probably with a positive metallicity-dependence suggested by hydrodynamic 
models.
   If we admit that the other parameters in this study are relatively better 
known than $\eta$, fixing the other parameters allows us to constrain the 
true value of $\eta$.
   In the metal-poor regime ($Z \lesssim 0.01$), if we assume that Galactic 
globular clusters are about 15 billion years old, the IMF slope 
$x \approx 1.35$, and $\sigma \approx 0.06$~\Msun, the models with 
$\eta \approx 0.3$ -- 0.5 fit metal-poor cluster data better.
   However, a larger value-models (e.g. $\eta \approx 0.5$ -- 0.7) fit 
relatively metal-rich globular cluster data (such as NGC\,6388) substantially 
better. 
   Similarly, under the same assumptions as the metal-poor case,
only high-$\eta$ (0.7 -- 1.0) models fit the elliptical galaxy data.

5. Mass dispersion on the HB is crucial to the UV flux. 
   A larger dispersion causes a stronger (weaker) UV flux when the HB stars
with the average mass are red (blue). 
   Because the effect of $\sigma$ is large, a population synthesis based on 
a non-realistic synthetic HB is likely to lead us to erroneous conclusions.
   We chose $\sigma = 0.06$~\Msun from globular cluster studies as our 
``standard'' dispersion.
   However, this value does not have a physical basis and is poorly known
for metal-rich systems. 

6. Even a significant change in the IMF slope, $x$, results in only a 
slight change in the resulting UV flux in comparison to optical flux. 
   A smaller value of $x$ causes a slightly stronger UV flux, because a model 
galaxy with a smaller $x$ sends more stars from the RGB to the HB at a fixed 
time. 
   However, there is little justification for an adoption of a much different 
IMF slope from the standard one, i.e. $x$ = 1.35.

\section{Conclusion}

   Based on this sensitivity study, we conclude as follows.
   Both the positive correlation between metallicity and UV-to-$V$ colors
among elliptical galaxies (\cite{b88}) and the negative correlation
between metallicity and UV-to-$V$ colors among globular clusters have been
explained.
   Old, metal-rich populations, such as giant elliptical galaxies, naturally
develop a UV upturn within a reasonable time scale ($\lesssim$ Hubble time)
without the necessity of the presence of young stars.
   Low-mass, core helium-burning (HB and evolved HB) stars are more
likely to be the main UV sources in the old stellar systems than any other
types of stars, as suggested before (e.g. \cite{gr90}; \cite{hdp92}; 
\cite{fd93}; \cite{bcf94}; \cite{dor95}; \cite{yado95}; \cite{bfdd97}). 
   For this reason, both metal-poor and metal-rich populations 
can develop a UV upturn if an arbitrarily large age may be assumed.

   Metal-poor models, such as globular clusters, are bluer in m(1500)-$V$ 
and m(2500)-$V$ than metal-rich models like elliptical galaxies as observations
suggest, mainly because of the opacity effect.
   But, their UV spectra are flat and fail to exhibit a strong UV upturn
because they lack very hot HB and post-HB stars.
   It takes too long ($>$ Hubble time) for a 
metal-poor population to develop a UV upturn of the observed strength if
$\eta \lesssim$ 0.7 for metal-poor stars as both observations and 
hydrodynamical models suggest.
   For this reason, no metal-poor stellar system has been observed to show 
a strong UV upturn (i.e. the continuous increase in the UV flux with decreasing
wavelength below 2,500 \AA) no matter how old it is. 
   Even if metal-poor models with a larger $\eta$ ($\eta \gtrsim$ 1.0,
an unrealistic assumption)
may reproduce the observed UV-to-$V$ colors of giant elliptical galaxies, 
the overall spectral shape in the UV would not match the observations very 
well, as will be quantitatively shown in our next paper (\cite{ydo97}).
   On the other hand, metal-rich model populations contain a sufficiently 
large number of UV-bright SBP stars, in addition to hot HB stars, to exhibit 
a UV upturn of the observed magnitude unless the input parameters assumed in 
this study are significantly wrong.
   Based on this argument, metal-poor stars cannot be the main cause of the 
UV upturn in giant elliptical galaxies that are obviously composite 
populations, as long as the majority of stars in giant elliptical galaxies 
are metal-rich.

   Two main effects drive a metal-rich population to develop a UV 
upturn of the observed magnitude earlier than a metal-poor population. 
   Firstly, metal-rich stars experience a larger amount of mass loss on the RGB
when Reimers' formula and a fixed $\eta$ are used. 
   This generates lower-mass HB stars at a fixed time.
   This is mainly because a higher opacity causes a larger stellar radius. 
   Reimers' mass loss formula (and our intuition as well) suggests 
that a smaller surface gravity caused by the larger radius results in a larger 
mass loss. 
   Secondly, even if some HB stars are not hot on the ZAHB, more metal-rich 
stars evolve to UV-bright stars more easily. 

   In conclusion, it is important to realize that the presence of the UV 
upturn in the spectrum of giant elliptical galaxies becomes neither 
extraordinary nor unexpected when proper treatments of various evolutionary 
stages are taken into account. 
   It is a very natural consequence of advanced stellar evolution that has 
not been known in detail until recently.
   While the presence of the UV upturn has been qualitatively understood, 
the detailed characteristics of the observed UV upturn also have to be 
reproduced in order to confirm the theoretical explanations.
   All these aspects will be investigated quantitatively in the 
following paper (\cite{ydo97}).

\acknowledgements

   We thank Richard Larson, Robert Zinn, Wayne Landsman, and Sally Heap 
for useful comments. 
   We are grateful to the anonymous referee for constructive criticisms
and clarifications on many points.
   This work was part of the Ph.D. study of S.Y. (\cite{yi96}) and was 
supported in part by NASA grants NAGW-3563 (S.Y. and A.O.), NAG5-1486 and 
NAG5-2469 (P.D.).
   Part of this work was performed while S.Y. held a National Research
Council-(NASA Goddard Space Flight Center) Research Associateship.

\clearpage

\begin{deluxetable}{rrrrrrrrrrrrrrr}
\footnotesize
\tablecaption{Mass loss estimates based on Reimers' formula ($\eta = 0.3$).}
\tablewidth{0pt}
\tablehead{
\colhead{t(Gyr)}&\colhead{ } &\colhead{ } &\colhead{ } &\colhead{ } &\colhead{ } &\colhead{ } & \colhead{$\Delta$$M$}& \colhead{ }&\colhead{ } &\colhead{ } &\colhead{ } &\colhead{ } & \colhead{ }\nl
                &\colhead{ } &\colhead{ } &\colhead{ } &\colhead{ } &\colhead{ } &\colhead{ } & \colhead{(\Msun)}& \colhead{ }&\colhead{ } &\colhead{ } &\colhead{ } &\colhead{ } & \colhead{ }\nl
\colhead{$Z$}& \colhead{0.0001}& \colhead{0.0004}& \colhead{0.001}& \colhead{0.004}& \colhead{0.01}& \colhead{0.02}& \colhead{0.04}& \colhead{0.06}& \colhead{0.1}& \colhead{0.01}& \colhead{0.02}& \colhead{0.04}& \colhead{0.06}& \colhead{0.1}\nl
\colhead{$Y$}& \colhead{0.23  }& \colhead{0.23  }& \colhead{0.23 }& \colhead{0.23 }& \colhead{0.25}& \colhead{0.27}& \colhead{0.31}& \colhead{0.35}& \colhead{0.43}& \colhead{0.26}& \colhead{0.29}& \colhead{0.35}& \colhead{0.41}& \colhead{0.53}
}
\startdata
 1& 0.032& 0.039& 0.045& 0.063& 0.057& 0.054& 0.048& 0.038& 0.017& 0.049& 0.049& 0.040& 0.029& 0.014\nl
 2& 0.039& 0.045& 0.052& 0.069& 0.065& 0.067& 0.066& 0.061& 0.044& 0.065& 0.066& 0.066& 0.068& 0.040\nl
 3& 0.045& 0.052& 0.058& 0.075& 0.073& 0.079& 0.084& 0.083& 0.063& 0.080& 0.083& 0.092& 0.092& 0.067\nl
 4& 0.052& 0.058& 0.065& 0.081& 0.081& 0.092& 0.102& 0.104& 0.078& 0.087& 0.091& 0.106& 0.101& 0.086\nl
 5& 0.058& 0.065& 0.071& 0.087& 0.089& 0.097& 0.113& 0.110& 0.094& 0.094& 0.097& 0.113& 0.111& 0.103\nl
 6& 0.065& 0.072& 0.078& 0.093& 0.096& 0.101& 0.119& 0.116& 0.109& 0.100& 0.102& 0.120& 0.120& 0.121\nl
 7& 0.070& 0.077& 0.084& 0.100& 0.103& 0.105& 0.124& 0.122& 0.124& 0.107& 0.108& 0.127& 0.129& 0.138\nl
 8& 0.072& 0.080& 0.087& 0.106& 0.110& 0.110& 0.130& 0.128& 0.136& 0.113& 0.113& 0.134& 0.138& 0.130\nl
 9& 0.075& 0.082& 0.090& 0.112& 0.117& 0.114& 0.135& 0.134& 0.140& 0.120& 0.119& 0.141& 0.143& 0.121\nl
10& 0.077& 0.085& 0.093& 0.115& 0.124& 0.119& 0.140& 0.140& 0.145& 0.126& 0.124& 0.148& 0.147& 0.112\nl
11& 0.079& 0.087& 0.095& 0.118& 0.131& 0.123& 0.146& 0.146& 0.149& 0.129& 0.130& 0.155& 0.152& 0.102\nl
12& 0.082& 0.090& 0.098& 0.120& 0.135& 0.128& 0.151& 0.151& 0.154& 0.133& 0.132& 0.158& 0.156& 0.093\nl
13& 0.084& 0.092& 0.100& 0.123& 0.138& 0.132& 0.157& 0.157& 0.158& 0.136& 0.135& 0.162& 0.160& 0.083\nl
14& 0.086& 0.094& 0.103& 0.125& 0.140& 0.136& 0.162& 0.161& 0.162& 0.139& 0.137& 0.165& 0.164& 0.074\nl
15& 0.088& 0.097& 0.105& 0.128& 0.143& 0.138& 0.167& 0.163& 0.166& 0.142& 0.139& 0.168& 0.168& 0.065\nl
16& 0.091& 0.099& 0.108& 0.131& 0.146& 0.141& 0.170& 0.166& 0.171& 0.146& 0.142& 0.172& 0.173& 0.056\nl
17& 0.093& 0.102& 0.111& 0.133& 0.149& 0.143& 0.173& 0.169& 0.175& 0.149& 0.144& 0.175& 0.177& 0.046\nl
18& 0.095& 0.104& 0.113& 0.136& 0.152& 0.146& 0.175& 0.172& 0.175& 0.152& 0.147& 0.178& 0.181& 0.037\nl
19& 0.097& 0.107& 0.116& 0.138& 0.155& 0.148& 0.178& 0.175& 0.174& 0.156& 0.149& 0.182& 0.181& 0.027\nl
20& 0.100& 0.109& 0.118& 0.141& 0.158& 0.151& 0.181& 0.178& 0.172& 0.159& 0.151& 0.185& 0.171& 0.024\nl
21& 0.102& 0.112& 0.121& 0.144& 0.161& 0.153& 0.183& 0.181& 0.171& 0.162& 0.154& 0.188& 0.161& 0.028\nl
22& 0.104& 0.114& 0.123& 0.146& 0.164& 0.155& 0.186& 0.183& 0.170& 0.165& 0.156& 0.191& 0.150& 0.033\nl
23& 0.106& 0.116& 0.126& 0.149& 0.167& 0.158& 0.189& 0.186& 0.168& 0.160& 0.159& 0.195& 0.140& 0.037\nl
24& 0.109& 0.119& 0.128& 0.151& 0.170& 0.160& 0.191& 0.189& 0.167& 0.147& 0.161& 0.198& 0.130& 0.042\nl
25& 0.111& 0.121& 0.131& 0.154& 0.173& 0.163& 0.194& 0.192& 0.166& 0.134& 0.163& 0.201& 0.120& 0.046\nl
\enddata
\end{deluxetable}
\clearpage

\begin{deluxetable}{rrrrrrrrrrrrrrr}
\footnotesize
\tablecaption{Mass loss estimates based on Reimers' formula ($\eta = 0.5$).}
\tablewidth{0pt}
\tablehead{
\colhead{t(Gyr)}&\colhead{ } &\colhead{ } &\colhead{ } &\colhead{ } &\colhead{ } &\colhead{ } & \colhead{$\Delta$$M$}& \colhead{ }&\colhead{ } &\colhead{ } &\colhead{ } &\colhead{ } & \colhead{ }\nl
                &\colhead{ } &\colhead{ } &\colhead{ } &\colhead{ } &\colhead{ } &\colhead{ } & \colhead{(\Msun)}& \colhead{ }&\colhead{ } &\colhead{ } &\colhead{ } &\colhead{ } & \colhead{ }\nl
\colhead{$Z$}& \colhead{0.0001}& \colhead{0.0004}& \colhead{0.001}& \colhead{0.004}& \colhead{0.01}& \colhead{0.02}& \colhead{0.04}& \colhead{0.06}& \colhead{0.1}& \colhead{0.01}& \colhead{0.02}& \colhead{0.04}& \colhead{0.06}& \colhead{0.1}\nl
\colhead{$Y$}& \colhead{0.23  }& \colhead{0.23  }& \colhead{0.23 }& \colhead{0.23 }& \colhead{0.25}& \colhead{0.27}& \colhead{0.31}& \colhead{0.35}& \colhead{0.43}& \colhead{0.26}& \colhead{0.29}& \colhead{0.35}& \colhead{0.41}& \colhead{0.53}
}
\startdata
 1& 0.054& 0.064& 0.075& 0.106& 0.093& 0.090& 0.078& 0.065& 0.029& 0.082& 0.082& 0.065& 0.048& 0.024\nl
 2& 0.065& 0.075& 0.086& 0.116& 0.108& 0.111& 0.109& 0.102& 0.073& 0.108& 0.110& 0.110& 0.113& 0.068\nl
 3& 0.076& 0.086& 0.097& 0.126& 0.122& 0.132& 0.140& 0.139& 0.105& 0.133& 0.138& 0.154& 0.154& 0.111\nl
 4& 0.087& 0.097& 0.108& 0.136& 0.136& 0.153& 0.171& 0.174& 0.131& 0.145& 0.152& 0.177& 0.170& 0.144\nl
 5& 0.097& 0.108& 0.119& 0.146& 0.149& 0.161& 0.189& 0.183& 0.156& 0.156& 0.161& 0.188& 0.185& 0.172\nl
 6& 0.108& 0.119& 0.130& 0.156& 0.161& 0.168& 0.198& 0.193& 0.182& 0.167& 0.170& 0.200& 0.200& 0.201\nl
 7& 0.117& 0.129& 0.140& 0.166& 0.172& 0.176& 0.207& 0.203& 0.207& 0.178& 0.179& 0.212& 0.215& 0.230\nl
 8& 0.122& 0.133& 0.145& 0.176& 0.184& 0.183& 0.216& 0.213& 0.227& 0.189& 0.189& 0.223& 0.231& 0.218\nl
 9& 0.126& 0.137& 0.149& 0.186& 0.195& 0.191& 0.225& 0.223& 0.234& 0.200& 0.198& 0.235& 0.239& 0.202\nl
10& 0.131& 0.141& 0.153& 0.189& 0.207& 0.198& 0.234& 0.233& 0.242& 0.211& 0.207& 0.246& 0.246& 0.186\nl
11& 0.136& 0.145& 0.156& 0.191& 0.219& 0.206& 0.243& 0.242& 0.249& 0.215& 0.216& 0.258& 0.253& 0.171\nl
12& 0.140& 0.149& 0.159& 0.193& 0.224& 0.213& 0.252& 0.252& 0.256& 0.219& 0.221& 0.263& 0.260& 0.155\nl
13& 0.145& 0.153& 0.162& 0.194& 0.229& 0.221& 0.261& 0.262& 0.263& 0.223& 0.225& 0.268& 0.267& 0.140\nl
14& 0.149& 0.157& 0.166& 0.196& 0.234& 0.227& 0.270& 0.267& 0.271& 0.227& 0.229& 0.273& 0.274& 0.124\nl
15& 0.154& 0.161& 0.169& 0.197& 0.239& 0.231& 0.278& 0.271& 0.278& 0.231& 0.233& 0.278& 0.281& 0.108\nl
16& 0.159& 0.165& 0.172& 0.199& 0.244& 0.235& 0.281& 0.275& 0.285& 0.235& 0.237& 0.283& 0.288& 0.093\nl
17& 0.163& 0.169& 0.176& 0.201& 0.249& 0.240& 0.284& 0.279& 0.292& 0.240& 0.241& 0.288& 0.295& 0.077\nl
18& 0.168& 0.173& 0.179& 0.202& 0.254& 0.244& 0.287& 0.283& 0.292& 0.244& 0.245& 0.293& 0.302& 0.062\nl
19& 0.172& 0.177& 0.182& 0.204& 0.258& 0.248& 0.289& 0.288& 0.288& 0.248& 0.249& 0.298& 0.302& 0.046\nl
20& 0.177& 0.181& 0.186& 0.206& 0.263& 0.252& 0.292& 0.292& 0.285& 0.252& 0.253& 0.303& 0.285& 0.039\nl
21& 0.182& 0.184& 0.189& 0.207& 0.268& 0.256& 0.295& 0.296& 0.282& 0.256& 0.257& 0.308& 0.268& 0.044\nl
22& 0.186& 0.189& 0.192& 0.209& 0.273& 0.260& 0.297& 0.300& 0.279& 0.260& 0.261& 0.313& 0.251& 0.049\nl
23& 0.191& 0.192& 0.196& 0.211& 0.278& 0.264& 0.300& 0.305& 0.276& 0.264& 0.265& 0.318& 0.234& 0.054\nl
24& 0.196& 0.196& 0.199& 0.212& 0.283& 0.268& 0.303& 0.309& 0.273& 0.268& 0.269& 0.323& 0.217& 0.058\nl
25& 0.200& 0.200& 0.202& 0.214& 0.288& 0.272& 0.306& 0.313& 0.269& 0.272& 0.273& 0.328& 0.200& 0.063\nl
\enddata
\end{deluxetable}
\clearpage

\begin{deluxetable}{rrrrrrrrrrrrrrr}
\footnotesize
\tablecaption{Mass loss estimates based on Reimers' formula ($\eta = 0.7$).}
\tablewidth{0pt}
\tablehead{
\colhead{t(Gyr)}&\colhead{ } &\colhead{ } &\colhead{ } &\colhead{ } &\colhead{ } &\colhead{ } & \colhead{$\Delta$$M$}& \colhead{ }&\colhead{ } &\colhead{ } &\colhead{ } &\colhead{ } & \colhead{ }\nl
                &\colhead{ } &\colhead{ } &\colhead{ } &\colhead{ } &\colhead{ } &\colhead{ } & \colhead{(\Msun)}& \colhead{ }&\colhead{ } &\colhead{ } &\colhead{ } &\colhead{ } & \colhead{ }\nl
\colhead{$Z$}& \colhead{0.0001}& \colhead{0.0004}& \colhead{0.001}& \colhead{0.004}& \colhead{0.01}& \colhead{0.02}& \colhead{0.04}& \colhead{0.06}& \colhead{0.1}& \colhead{0.01}& \colhead{0.02}& \colhead{0.04}& \colhead{0.06}& \colhead{0.1}\nl
\colhead{$Y$}& \colhead{0.23  }& \colhead{0.23  }& \colhead{0.23 }& \colhead{0.23 }& \colhead{0.25}& \colhead{0.27}& \colhead{0.31}& \colhead{0.35}& \colhead{0.43}& \colhead{0.26}& \colhead{0.29}& \colhead{0.35}& \colhead{0.41}& \colhead{0.53}
}
\startdata
 1& 0.077& 0.090& 0.105& 0.149& 0.131& 0.125& 0.110& 0.090& 0.040& 0.115& 0.115& 0.092& 0.067& 0.033\nl
 2& 0.092& 0.105& 0.120& 0.163& 0.151& 0.155& 0.153& 0.142& 0.103& 0.151& 0.153& 0.154& 0.157& 0.094\nl
 3& 0.107& 0.121& 0.135& 0.176& 0.170& 0.185& 0.196& 0.193& 0.148& 0.187& 0.192& 0.216& 0.215& 0.156\nl
 4& 0.122& 0.136& 0.151& 0.191& 0.189& 0.215& 0.239& 0.243& 0.183& 0.203& 0.212& 0.247& 0.237& 0.201\nl
 5& 0.137& 0.151& 0.166& 0.204& 0.208& 0.226& 0.264& 0.257& 0.218& 0.219& 0.225& 0.263& 0.259& 0.242\nl
 6& 0.152& 0.167& 0.181& 0.218& 0.224& 0.236& 0.277& 0.271& 0.254& 0.234& 0.238& 0.280& 0.280& 0.282\nl
 7& 0.163& 0.180& 0.196& 0.232& 0.241& 0.247& 0.290& 0.284& 0.289& 0.249& 0.251& 0.296& 0.302& 0.322\nl
 8& 0.166& 0.182& 0.199& 0.246& 0.257& 0.257& 0.302& 0.298& 0.315& 0.265& 0.264& 0.312& 0.323& 0.305\nl
 9& 0.168& 0.184& 0.202& 0.260& 0.273& 0.267& 0.315& 0.312& 0.320& 0.280& 0.277& 0.328& 0.330& 0.283\nl
10& 0.171& 0.186& 0.204& 0.263& 0.289& 0.278& 0.328& 0.326& 0.326& 0.295& 0.290& 0.345& 0.333& 0.261\nl
11& 0.174& 0.189& 0.206& 0.262& 0.306& 0.288& 0.340& 0.340& 0.331& 0.298& 0.303& 0.361& 0.336& 0.239\nl
12& 0.176& 0.191& 0.207& 0.262& 0.309& 0.299& 0.353& 0.354& 0.336& 0.301& 0.306& 0.361& 0.338& 0.217\nl
13& 0.179& 0.193& 0.209& 0.261& 0.311& 0.309& 0.365& 0.367& 0.341& 0.303& 0.307& 0.360& 0.341& 0.196\nl
14& 0.182& 0.195& 0.211& 0.261& 0.312& 0.316& 0.378& 0.368& 0.346& 0.306& 0.309& 0.360& 0.344& 0.174\nl
15& 0.184& 0.197& 0.212& 0.260& 0.314& 0.318& 0.389& 0.368& 0.351& 0.309& 0.310& 0.359& 0.347& 0.152\nl
16& 0.187& 0.199& 0.214& 0.260& 0.316& 0.319& 0.387& 0.367& 0.356& 0.312& 0.312& 0.359& 0.350& 0.130\nl
17& 0.190& 0.202& 0.215& 0.259& 0.317& 0.320& 0.385& 0.366& 0.361& 0.315& 0.314& 0.358& 0.352& 0.108\nl
18& 0.192& 0.204& 0.217& 0.259& 0.319& 0.321& 0.384& 0.366& 0.356& 0.317& 0.315& 0.358& 0.355& 0.086\nl
19& 0.195& 0.206& 0.219& 0.258& 0.320& 0.322& 0.382& 0.365& 0.349& 0.320& 0.317& 0.357& 0.352& 0.065\nl
20& 0.198& 0.208& 0.220& 0.258& 0.322& 0.323& 0.380& 0.365& 0.342& 0.323& 0.318& 0.357& 0.334& 0.054\nl
21& 0.200& 0.210& 0.222& 0.257& 0.324& 0.325& 0.378& 0.364& 0.335& 0.326& 0.320& 0.356& 0.316& 0.059\nl
22& 0.203& 0.212& 0.223& 0.257& 0.325& 0.326& 0.377& 0.363& 0.327& 0.329& 0.322& 0.356& 0.298& 0.064\nl
23& 0.206& 0.215& 0.225& 0.256& 0.327& 0.327& 0.375& 0.363& 0.320& 0.318& 0.323& 0.355& 0.280& 0.068\nl
24& 0.208& 0.217& 0.226& 0.256& 0.328& 0.328& 0.373& 0.362& 0.313& 0.295& 0.325& 0.355& 0.262& 0.073\nl
25& 0.211& 0.219& 0.228& 0.255& 0.330& 0.329& 0.371& 0.361& 0.305& 0.272& 0.326& 0.354& 0.244& 0.078\nl
\enddata
\end{deluxetable}
\clearpage

\begin{deluxetable}{rrrrrrrrrrrrrrr}
\footnotesize
\tablecaption{Mass loss estimates based on Reimers' formula ($\eta = 1.0$).}
\tablewidth{0pt}
\tablehead{
\colhead{t(Gyr)}&\colhead{ } &\colhead{ } &\colhead{ } &\colhead{ } &\colhead{ } &\colhead{ } & \colhead{$\Delta$$M$}& \colhead{ }&\colhead{ } &\colhead{ } &\colhead{ } &\colhead{ } & \colhead{ }\nl
                &\colhead{ } &\colhead{ } &\colhead{ } &\colhead{ } &\colhead{ } &\colhead{ } & \colhead{(\Msun)}& \colhead{ }&\colhead{ } &\colhead{ } &\colhead{ } &\colhead{ } & \colhead{ }\nl
\colhead{$Z$}& \colhead{0.0001}& \colhead{0.0004}& \colhead{0.001}& \colhead{0.004}& \colhead{0.01}& \colhead{0.02}& \colhead{0.04}& \colhead{0.06}& \colhead{0.1}& \colhead{0.01}& \colhead{0.02}& \colhead{0.04}& \colhead{0.06}& \colhead{0.1}\nl
\colhead{$Y$}& \colhead{0.23  }& \colhead{0.23  }& \colhead{0.23 }& \colhead{0.23 }& \colhead{0.25}& \colhead{0.27}& \colhead{0.31}& \colhead{0.35}& \colhead{0.43}& \colhead{0.26}& \colhead{0.29}& \colhead{0.35}& \colhead{0.41}& \colhead{0.53}
}
\startdata
 1& 0.109& 0.129& 0.150& 0.212& 0.188& 0.179& 0.157& 0.130& 0.057& 0.164& 0.164& 0.132& 0.095& 0.047\nl
 2& 0.131& 0.151& 0.172& 0.232& 0.216& 0.222& 0.219& 0.203& 0.146& 0.216& 0.219& 0.220& 0.225& 0.134\nl
 3& 0.152& 0.172& 0.194& 0.252& 0.243& 0.265& 0.280& 0.277& 0.211& 0.267& 0.275& 0.309& 0.308& 0.222\nl
 4& 0.174& 0.194& 0.215& 0.272& 0.271& 0.307& 0.342& 0.347& 0.261& 0.291& 0.303& 0.354& 0.338& 0.271\nl
 5& 0.196& 0.216& 0.237& 0.292& 0.298& 0.322& 0.376& 0.367& 0.312& 0.313& 0.321& 0.377& 0.369& 0.307\nl
 6& 0.217& 0.238& 0.259& 0.312& 0.321& 0.337& 0.391& 0.387& 0.362& 0.335& 0.340& 0.400& 0.400& 0.343\nl
 7& 0.232& 0.257& 0.280& 0.332& 0.344& 0.352& 0.407& 0.406& 0.413& 0.356& 0.359& 0.423& 0.431& 0.379\nl
 8& 0.232& 0.256& 0.281& 0.352& 0.367& 0.367& 0.422& 0.426& 0.447& 0.378& 0.377& 0.446& 0.461& 0.358\nl
 9& 0.232& 0.255& 0.282& 0.372& 0.390& 0.382& 0.438& 0.446& 0.440& 0.400& 0.396& 0.470& 0.461& 0.334\nl
10& 0.232& 0.254& 0.281& 0.373& 0.414& 0.396& 0.453& 0.466& 0.434& 0.422& 0.414& 0.493& 0.453& 0.309\nl
11& 0.233& 0.254& 0.280& 0.369& 0.437& 0.411& 0.468& 0.485& 0.427& 0.417& 0.433& 0.516& 0.445& 0.285\nl
12& 0.233& 0.253& 0.278& 0.365& 0.435& 0.426& 0.484& 0.505& 0.421& 0.411& 0.431& 0.507& 0.437& 0.260\nl
13& 0.233& 0.252& 0.276& 0.361& 0.429& 0.441& 0.499& 0.525& 0.414& 0.406& 0.427& 0.497& 0.428& 0.236\nl
14& 0.233& 0.252& 0.275& 0.357& 0.424& 0.449& 0.515& 0.520& 0.407& 0.401& 0.424& 0.487& 0.420& 0.211\nl
15& 0.233& 0.251& 0.273& 0.352& 0.418& 0.444& 0.527& 0.511& 0.401& 0.396& 0.420& 0.478& 0.412& 0.187\nl
16& 0.233& 0.250& 0.271& 0.348& 0.412& 0.439& 0.519& 0.503& 0.394& 0.390& 0.417& 0.468& 0.403& 0.162\nl
17& 0.234& 0.249& 0.270& 0.344& 0.406& 0.434& 0.512& 0.494& 0.387& 0.385& 0.413& 0.459& 0.395& 0.138\nl
18& 0.234& 0.249& 0.268& 0.340& 0.400& 0.429& 0.504& 0.486& 0.380& 0.380& 0.410& 0.449& 0.387& 0.114\nl
19& 0.234& 0.248& 0.267& 0.336& 0.395& 0.424& 0.496& 0.477& 0.373& 0.374& 0.406& 0.440& 0.377& 0.089\nl
20& 0.234& 0.247& 0.265& 0.332& 0.389& 0.419& 0.488& 0.469& 0.365& 0.369& 0.403& 0.430& 0.362& 0.077\nl
21& 0.234& 0.247& 0.263& 0.328& 0.383& 0.414& 0.481& 0.460& 0.358& 0.364& 0.399& 0.421& 0.347& 0.081\nl
22& 0.235& 0.246& 0.262& 0.324& 0.377& 0.410& 0.473& 0.452& 0.350& 0.359& 0.396& 0.411& 0.332& 0.086\nl
23& 0.235& 0.245& 0.260& 0.320& 0.372& 0.405& 0.465& 0.443& 0.343& 0.346& 0.392& 0.402& 0.317& 0.090\nl
24& 0.235& 0.244& 0.259& 0.316& 0.366& 0.400& 0.458& 0.435& 0.336& 0.328& 0.389& 0.392& 0.303& 0.094\nl
25& 0.235& 0.244& 0.257& 0.312& 0.360& 0.395& 0.450& 0.426& 0.328& 0.309& 0.385& 0.382& 0.288& 0.099\nl
\enddata
\end{deluxetable}

\clearpage

\begin{table*}[t]
\caption{Best fitting $\eta$ estimates for metal-poor stars.}
\begin{center}
\begin{tabular}{rrcrrc}
\tableline
\tableline
$Z$   & $M_{RR}$ \tablenotemark{a} & Age(Gyr) \tablenotemark{b} & $M_{RG}$ \tablenotemark{c} & $\Delta M$ \tablenotemark{d} & $\eta$ \\
\tableline
0.0001 & 0.760  & 12   & 0.852 &  0.092 & 0.3 -- 0.4\\
0.0001 & 0.760  & 14   & 0.816 &  0.056 & $<$ 0.3  \\
0.0001 & 0.760  & 16   & 0.788 &  0.028 & $<$ 0.3  \\

0.0004 & 0.710  & 12   & 0.862 &  0.152 &    0.5   \\
0.0004 & 0.710  & 14   & 0.826 &  0.116 & 0.3 -- 0.5\\
0.0004 & 0.710  & 16   & 0.796 &  0.086 & $<$ 0.3  \\
\tableline
\tableline
\end{tabular}
\end{center}
\tablenotetext{a}{Evolutionary mass of RR Lyrae type-d variables from Yi et al. (1993). }
\tablenotetext{b}{Because of uncertainties, three ages have been chosen for the clusters, M15 and M3.}
\tablenotetext{c}{Mass of red giants at the tip of the RGB, adopted from the Yale
Isochrones 1996.}
\tablenotetext{d}{\DM $\equiv M_{RG} - M_{RR}$}
\end{table*}

\clearpage

\begin{table*}[t]
\caption{The $\eta$ suggested by hydrodynamical models.}
\begin{center}
\begin{tabular}{rc}
\tableline
\tableline
$Z$  & suggested $\eta$ \\
\tableline
0.0002& 0.17 -- 0.25  \\
0.002 & 0.22 -- 0.33  \\
0.006 & 0.33 -- 0.50  \\
0.02  &     1.0      \\
\tableline
\tableline
\end{tabular}
\end{center}
\end{table*}

{}

\clearpage

\begin{figure}
\plotone{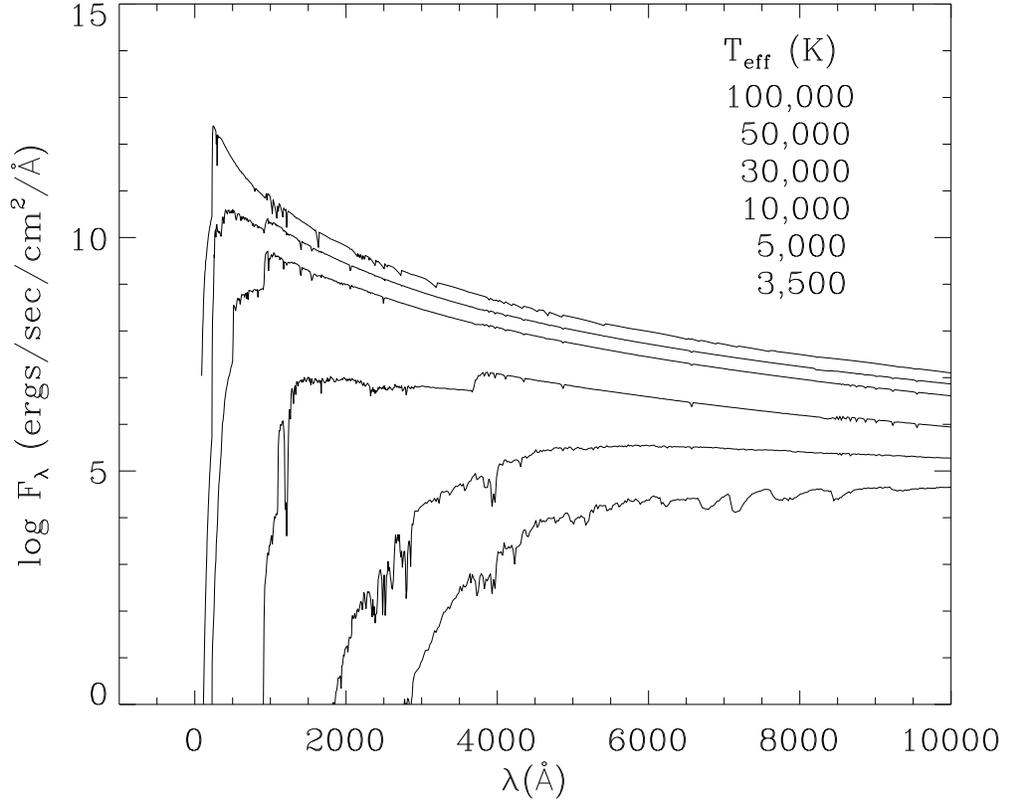}
\caption{
Variation of SED as a function of temperature for $Z =$ \Zsun and
\logg = 5.0. From the top, \Teff = 100,000, 50,000, 
30,000, 10,000, 5,000, \& 3,500 K. Model spectra of \Teff $>$ 50,000 have been 
constructed using Hubeny's spectral synthesis code while others are all from 
the Kurucz spectral library.
}\label{Figure 1}
\end{figure}

\clearpage

\begin{figure}
\plotone{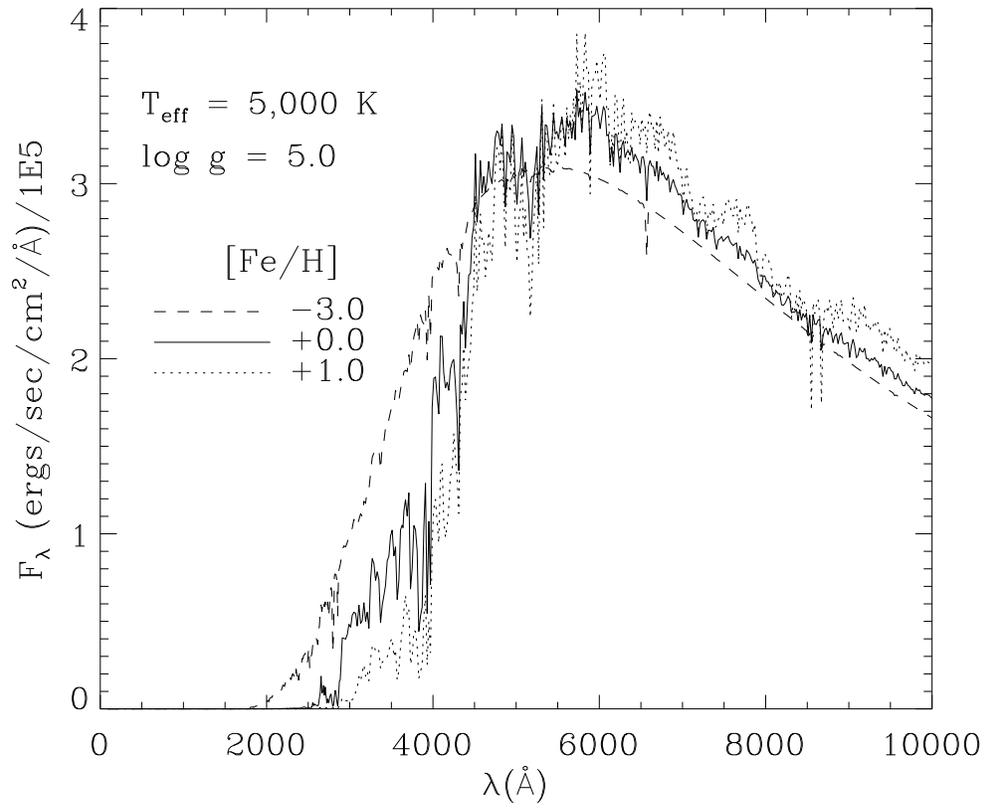}
\caption{
Variation of SED as a function of metallicity for \feh = $-$3, 0, and
$+$1, all for \Teff = 5,000 K and \logg = 5.0.
}\label{Figure 2}
\end{figure}

\clearpage

\begin{figure}
\plotone{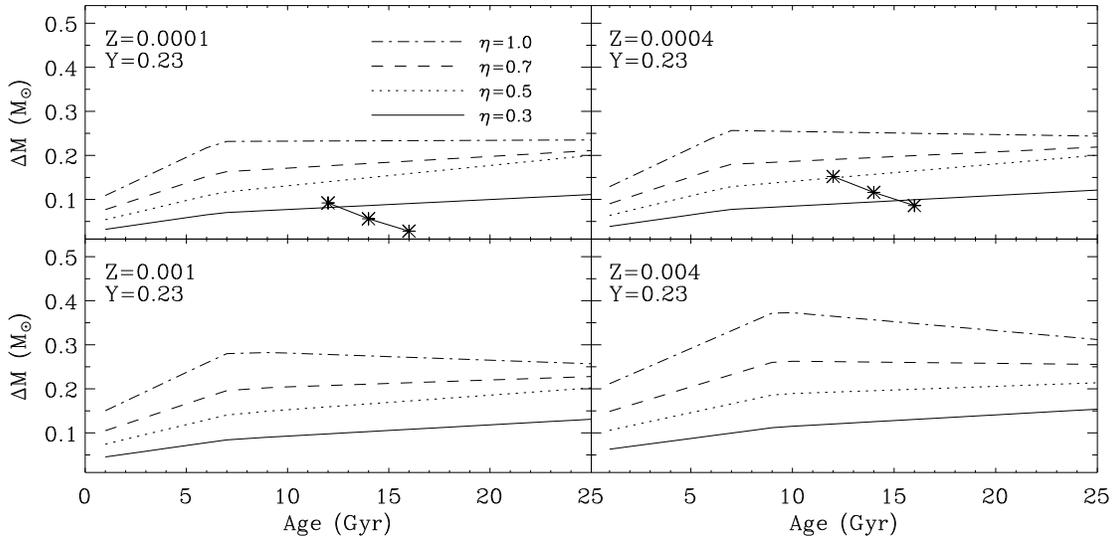}
\caption{
Mass loss estimates for metal-poor stars based on Reimers' formula with a
variety of efficiency parameters $\eta$. Estimation has been made using
stellar evolutionary tracks that were used to construct the Yale Isochrones
1996 .
In the top two panels, the lines with asterisks are from the measured mass of 
the RR Lyrae stars, $M_{RR}$, in the globular clusters assuming the age range
of 12 -- 16~Gyrs as shown in Table 5. The $M_{RR}$ for $Z = 0.0001$ and
$Z = 0.0004$ are best reproduced by $\eta \leq$ 0.3 and 
$\eta$ $\approx$ 0.3 -- 0.5, respectively.
}\label{Figure 3}
\end{figure}

\clearpage

\begin{figure}
\epsscale{0.7}
\plotone{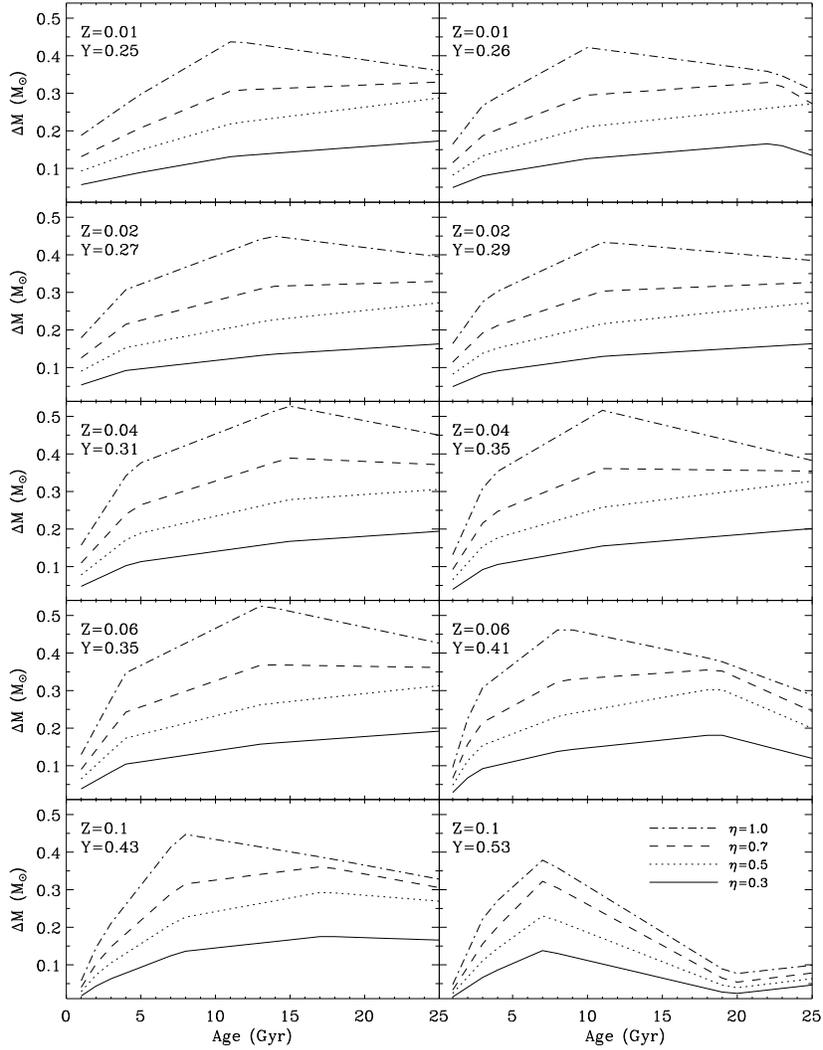}
\caption{
Same as Figure 3 but for metal-rich stars. The left and right panels are
for \DYDZ = 2 and 3, respectively. Mass loss as a function of age cannot 
increase indefinitely as $\eta$ increases because the mass of the red giants
and their core mass also depend on age and metallicity. After a critical age,
mass loss decreases for this reason.
}\label{Figure 4}
\end{figure}

\clearpage

\begin{figure}
\plotone{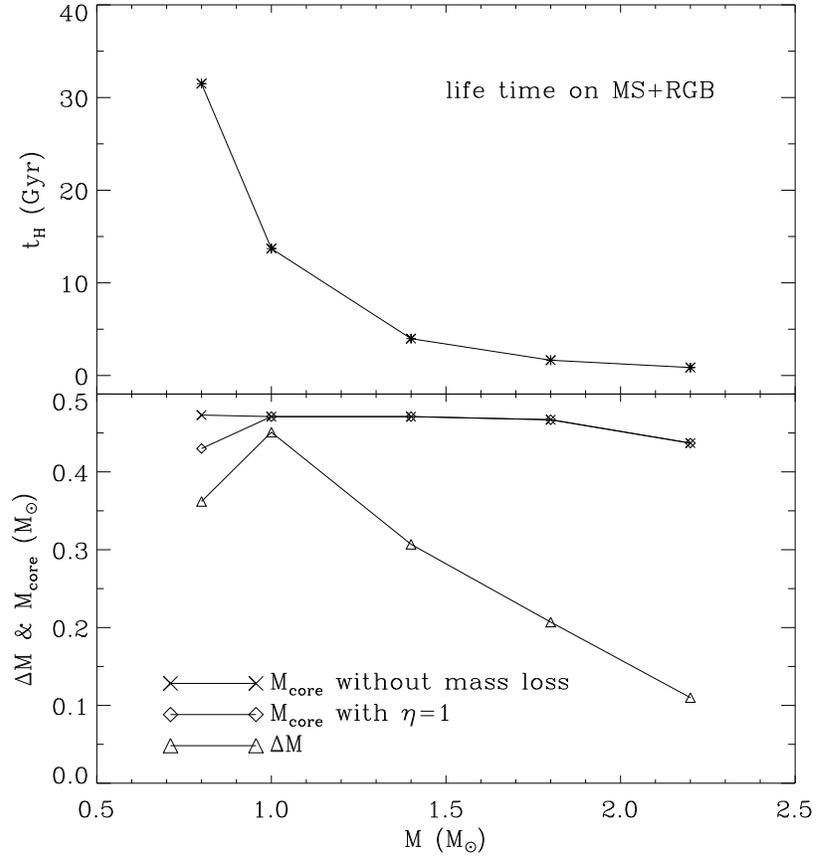}
\caption{
Lifetime on the MS and RGB, $t_{H}$ (upper panel), the core mass at the tip 
of the RGB, \Mcore, and the mass loss, \DM (lower panel), as a function of 
initial mass $M$ for $Z$ = \Zsun. \DM increases rapidly as $M$ decreases 
because $t_{H}$ increases as $M$ decreases. Below a certain mass, 
$M$ $\approx$ 1.0 in case of $Z$ = \Zsun, however, \DM decreases as $M$ 
decreases because \Mcore grows too fast to allow more mass loss to occur. 
A different mass loss efficiency $\eta$, therefore, leads to a different 
final core mass.
}\label{Figure 5}
\end{figure}

\clearpage

\begin{figure}
\plotone{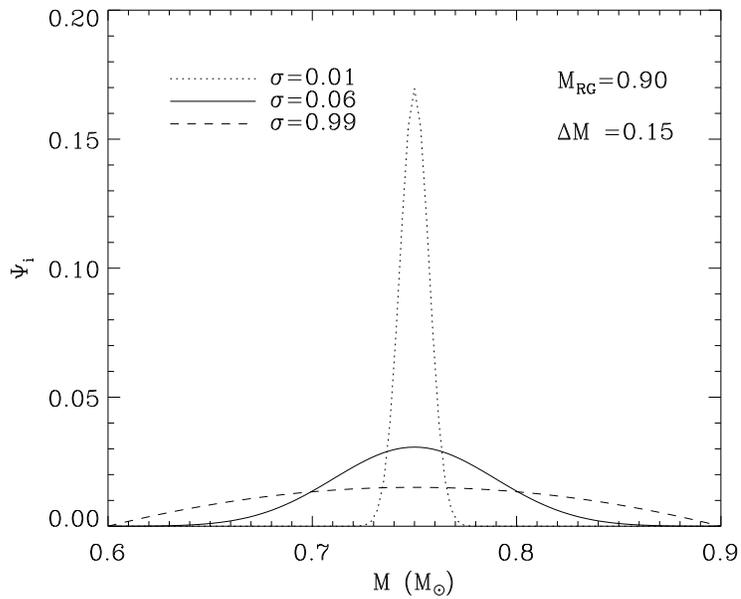}
\caption{
Mass distribution functions assumed in this population synthesis. The model
with $\sigma$ = 0.06~\Msun reproduces the width of the HB of Galactic globular
clusters (Lee et al. 1990). The models with $\sigma$ = 0.01 and 
0.99 are the two extreme assumptions that represent a pseudo-delta 
function and a pseudo-constant function for mass distribution. 
}\label{Figure 6}
\end{figure}

\clearpage

\begin{figure}
\plotone{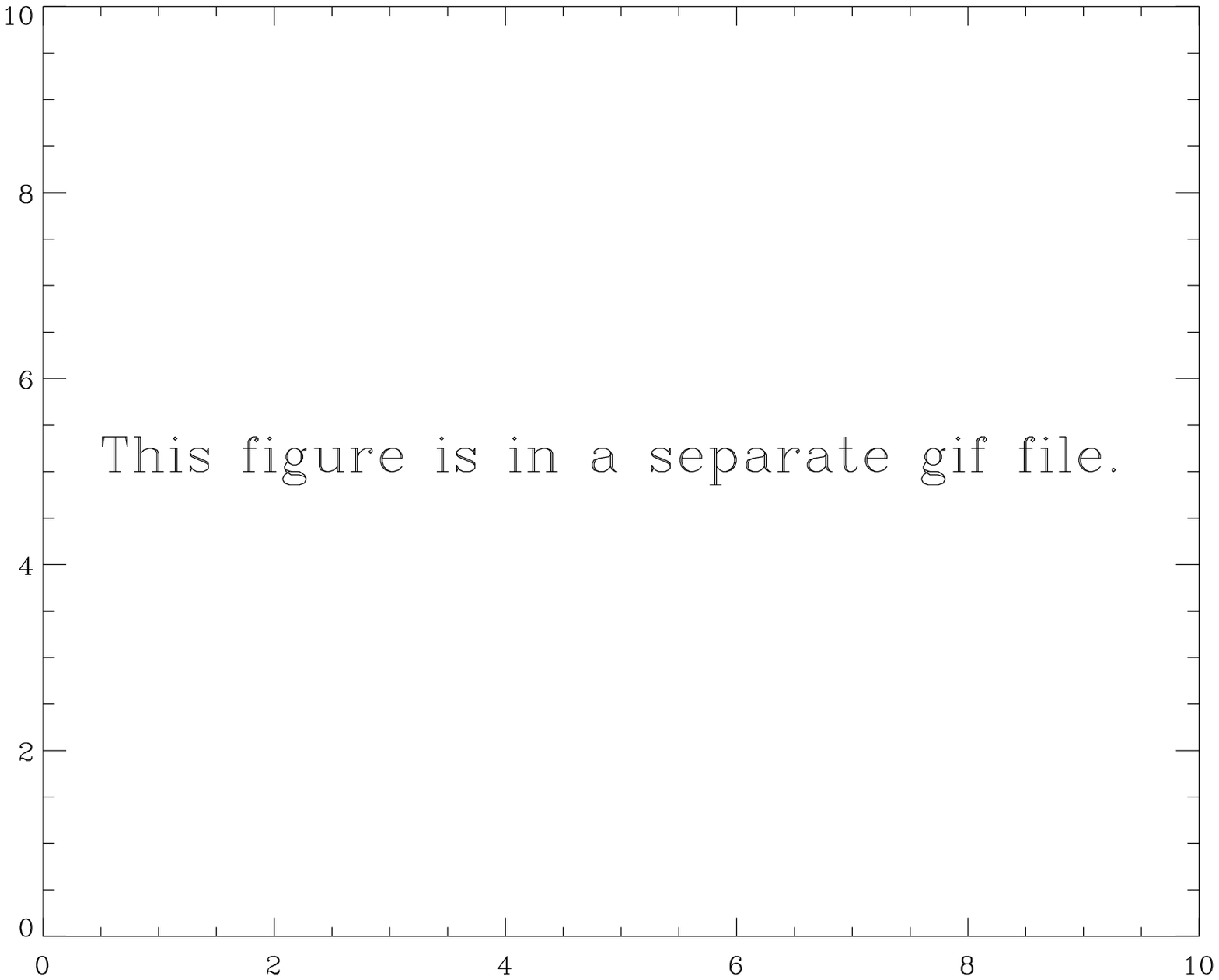}
\caption{
Model CMD for low metallicity as a function of age and metallicity. The 
synthetic HBs are based on the assumption of ($\eta$, $x$, $\sigma$) = 
(1.0, 1.35, 0.06).
Note that the HB becomes hotter as age increases.
A larger circle denotes a larger number of stars, where the total initial
mass of the model galaxy is $10^{12}$~\Msun.
It is important to note that these models are based on $\eta = 1.0$.
We think that this value is adequate only for metal-rich stars and is much 
higher than the estimated value ($\approx 0.3$ -- 0.5) for metal-poor stars.
As a result, the 15~Gyr old $Z = 0.004$ model with $\eta$ = 1.0 (bottom right
panel) contains too many blue HB stars compared to 47 Tuc. 
Nevertheless, $\eta$ has been kept the same for all models of different ages
and metallicities in Figures 7 -- 12 in order to see only the effect of age and
metallicity. The brightest PAGB track (in 15~Gyr-old models) is the Kiel 
group 0.598~\Msun model, and the second and third are their 0.565 and 
0.546~\Msun models, respectively. The fourth brightest horizontally crossing
track is for the stars that do not have a sufficiently large mass to experience
helium core flash (Sweigart et al. 1974; see text).
}\label{Figure 7}
\end{figure}

\clearpage

\begin{figure}
\epsscale{0.7}
\plotone{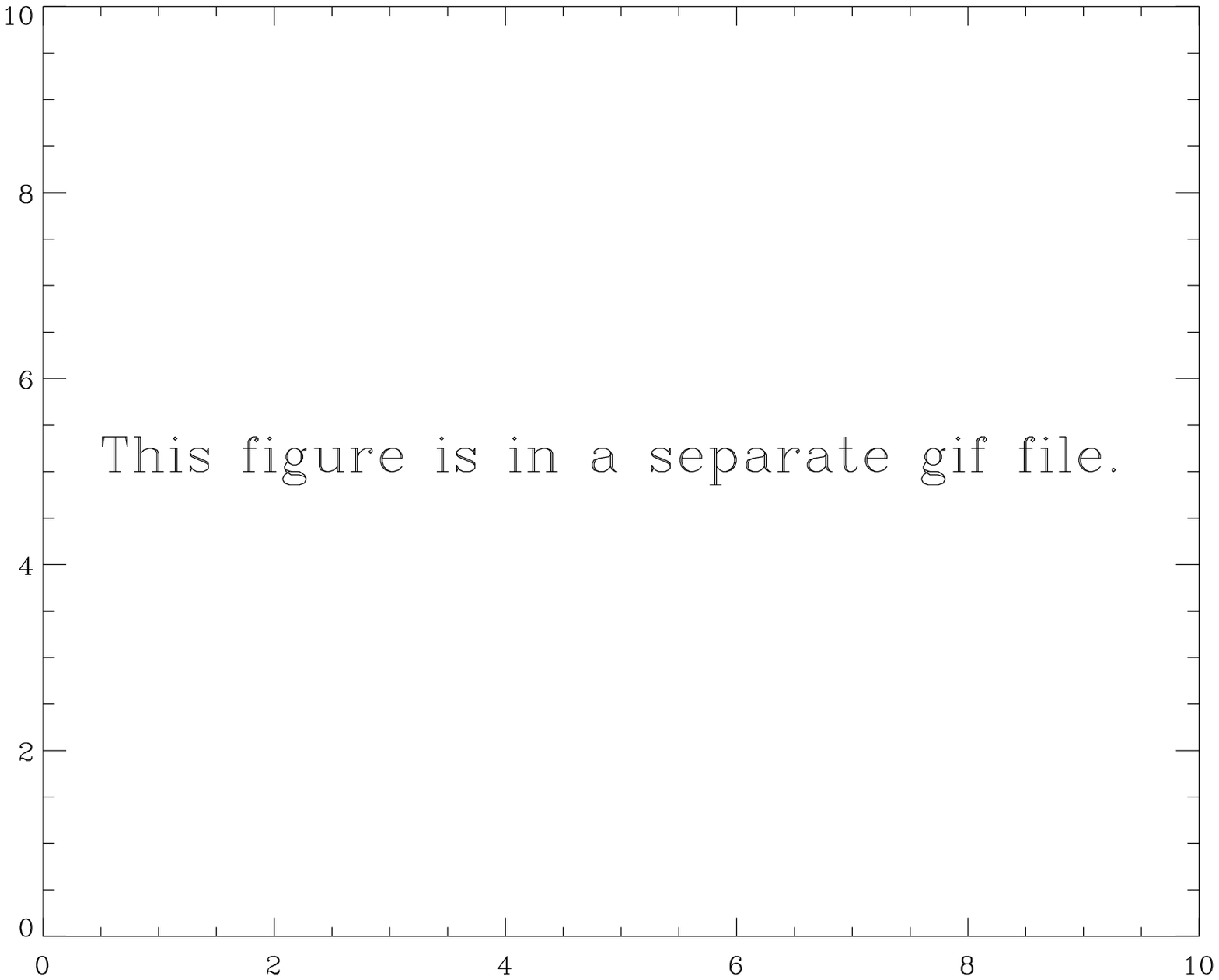}
\caption{
Same as Figure 7, but for metal-rich populations of $Z \geq$ 0.01 and 
\DYDZ = 2. Note that metal-rich models have many UV-bright (slow blue phase)
stars.
}\label{Figure 8}
\end{figure}

\clearpage

\begin{figure}
\epsscale{0.7}
\plotone{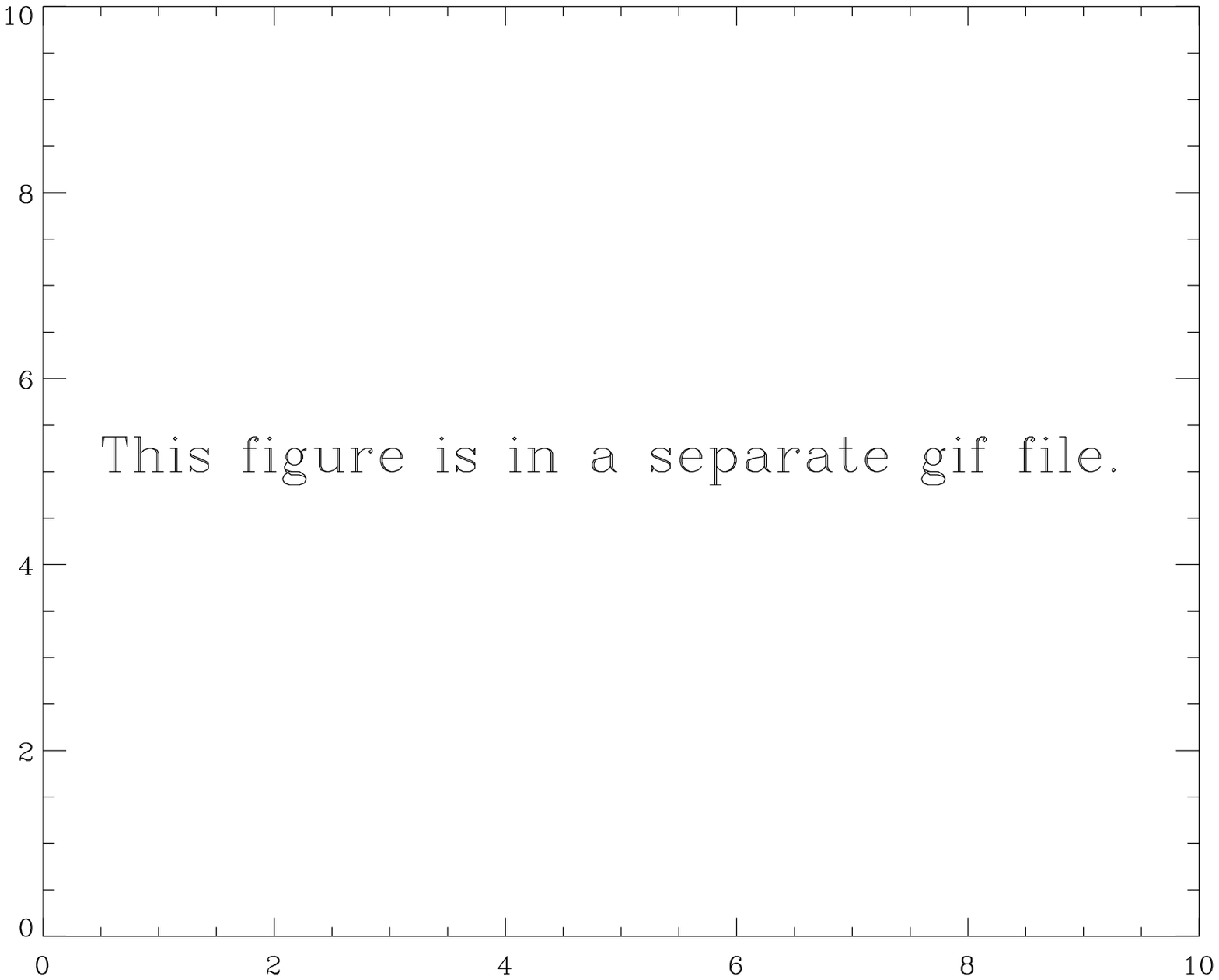}
\caption{
Same as Figure 7, but for metal-rich populations of $Z \geq$ 0.01 and 
\DYDZ = 3. The UV-bright phase, the SBP, is even more
conspicuous than the case of \DYDZ = 2 in Figure 8.
}\label{Figure 9}
\end{figure}

\clearpage

\begin{figure}
\plotone{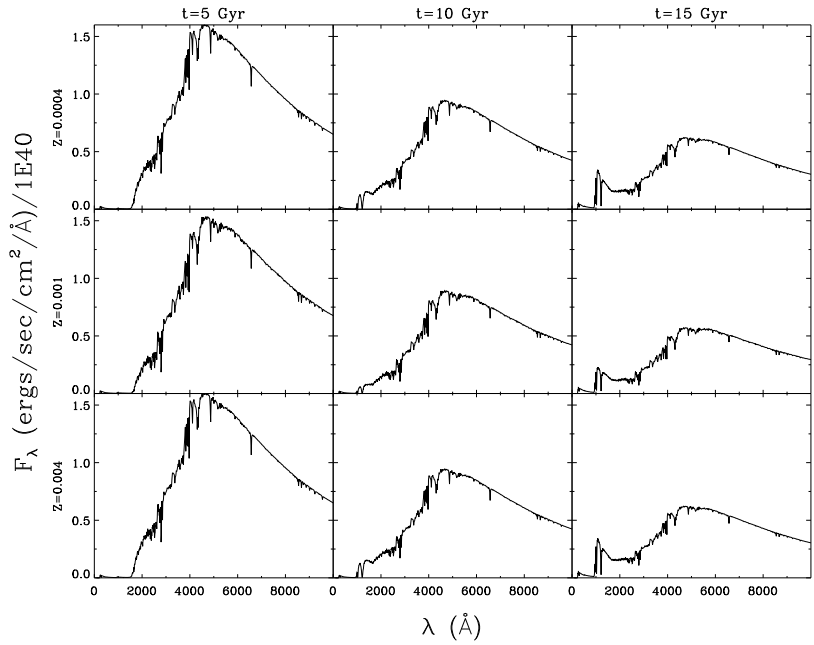}
\caption{
Model SED for low metallicity as a function of age and metallicity. These 
models are based on the model CMDs shown in Figure 7. 
A noticeable ``UV upturn'' appears at a large age. 
Once again, keep in mind that the metal-poor
models overproduce UV flux almost certainly because $\eta = 1.0$, a
factor of 2 -- 3 overestimation for metal-poor stars, has been used in this
figure. The y-axis is in an arbitrary scale.
}\label{Figure 10}
\end{figure}

\clearpage

\begin{figure}
\epsscale{0.7}
\plotone{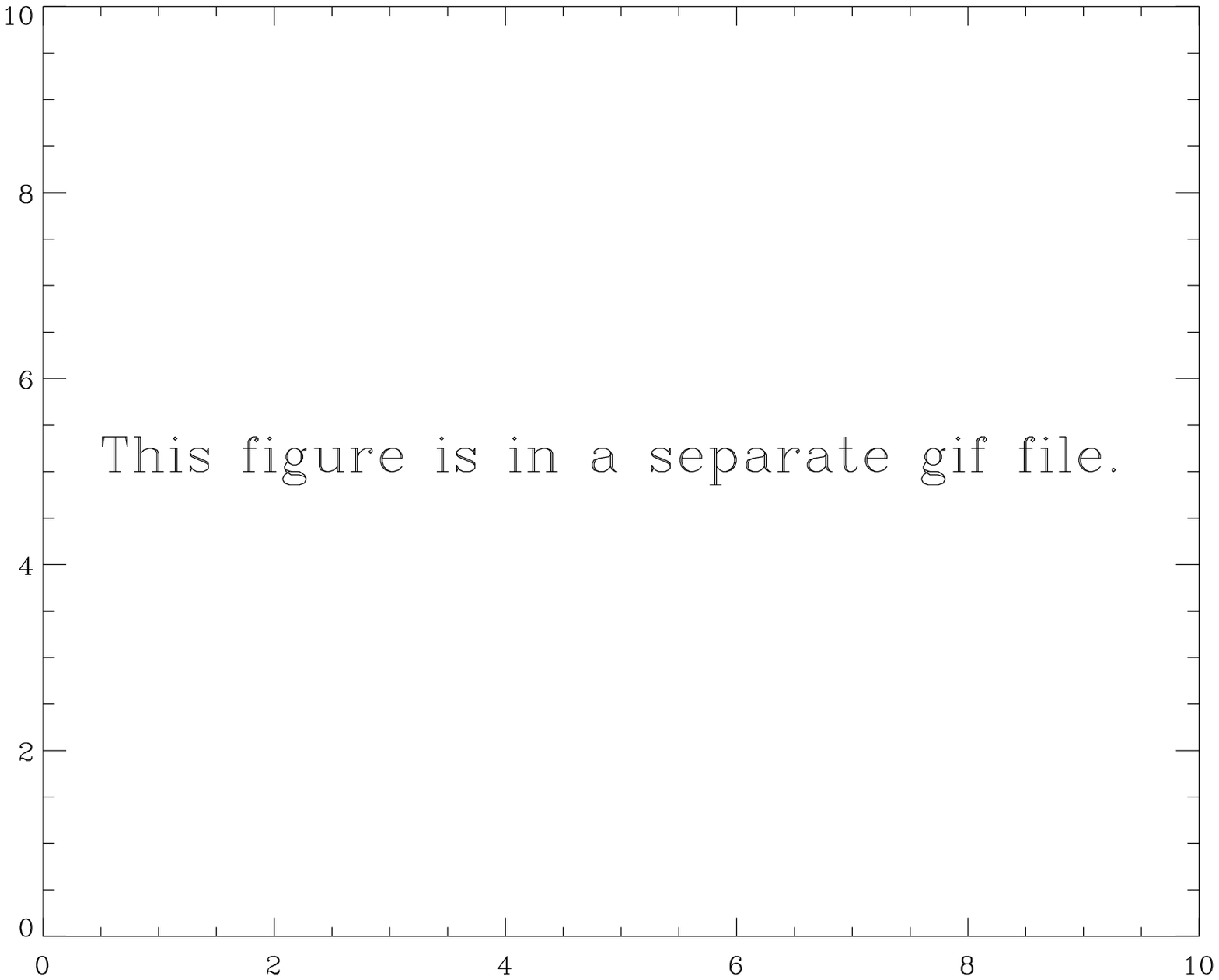}
\caption{
Same as Figure 10, but for $Z \geq$ 0.01 and \DYDZ = 2. The far-UV flux
relative to the near-UV flux is generally higher than the metal-poor models
in Figure 10.
}\label{Figure 11}
\end{figure}

\clearpage

\begin{figure}
\epsscale{0.7}
\plotone{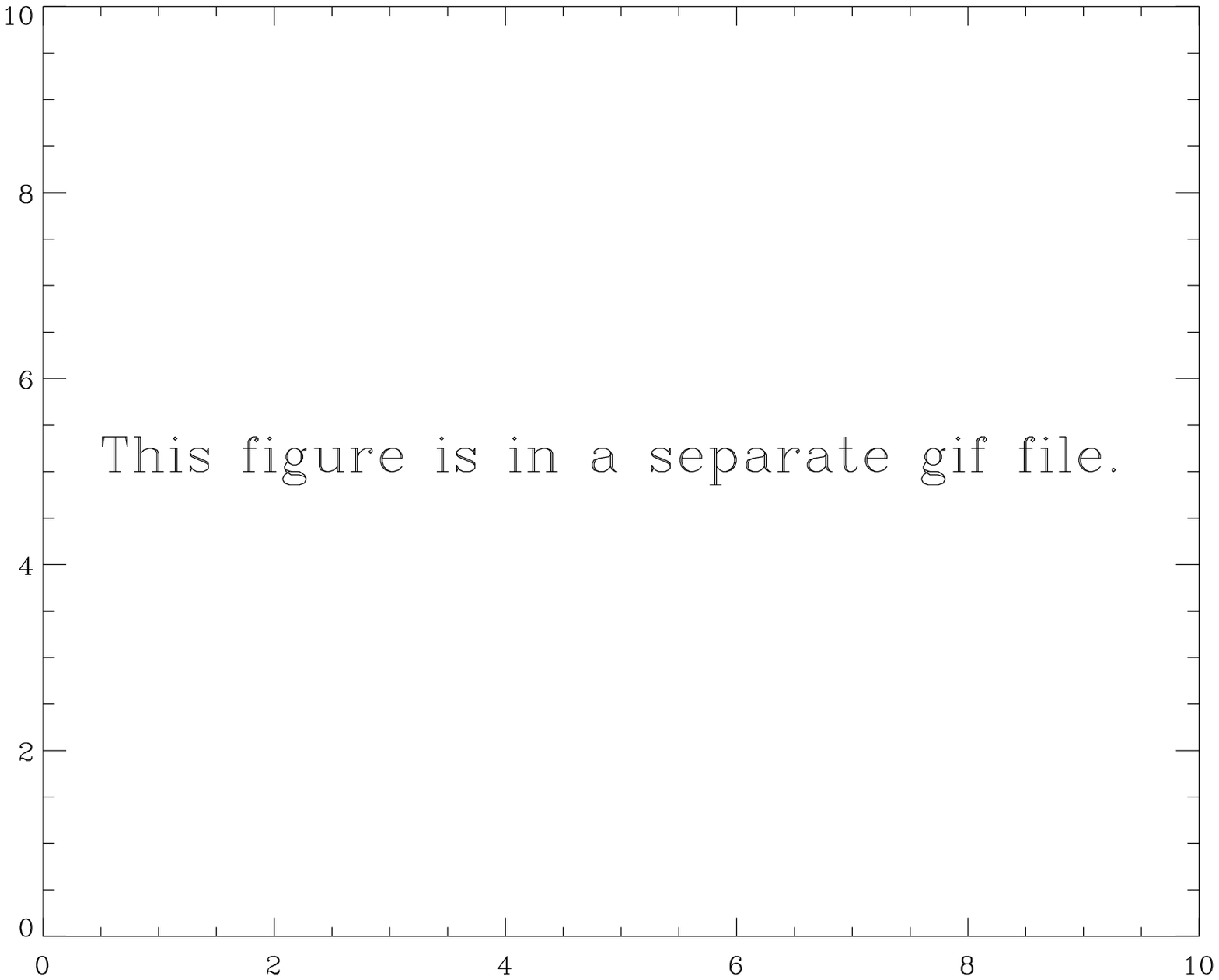}
\caption{
Same as Figure 11, but for \DYDZ = 3. A strong UV flux develops more quickly
when \DYDZ is higher because the UV-bright shell helium-burning phase (the
SBP) is more significant when helium abundance is higher (\cite{dro93}; YDK).
}\label{Figure 12}
\end{figure}

\clearpage

\begin{figure}
\plotone{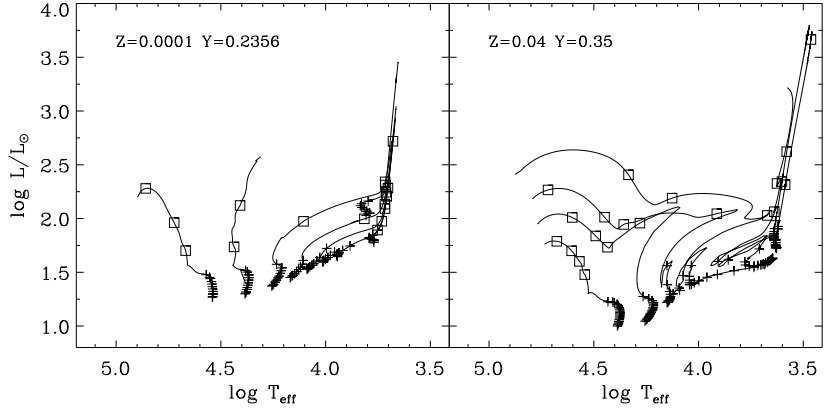}
\caption{
The definitions of central and shell helium-burning stages. The crosses and
squares are for central and shell helium-burning stages, respectively.
All symbols denote time interval of 10 Myr.
}\label{Figure 13}
\end{figure}

\clearpage

\begin{figure}
\epsscale{0.7}
\plotone{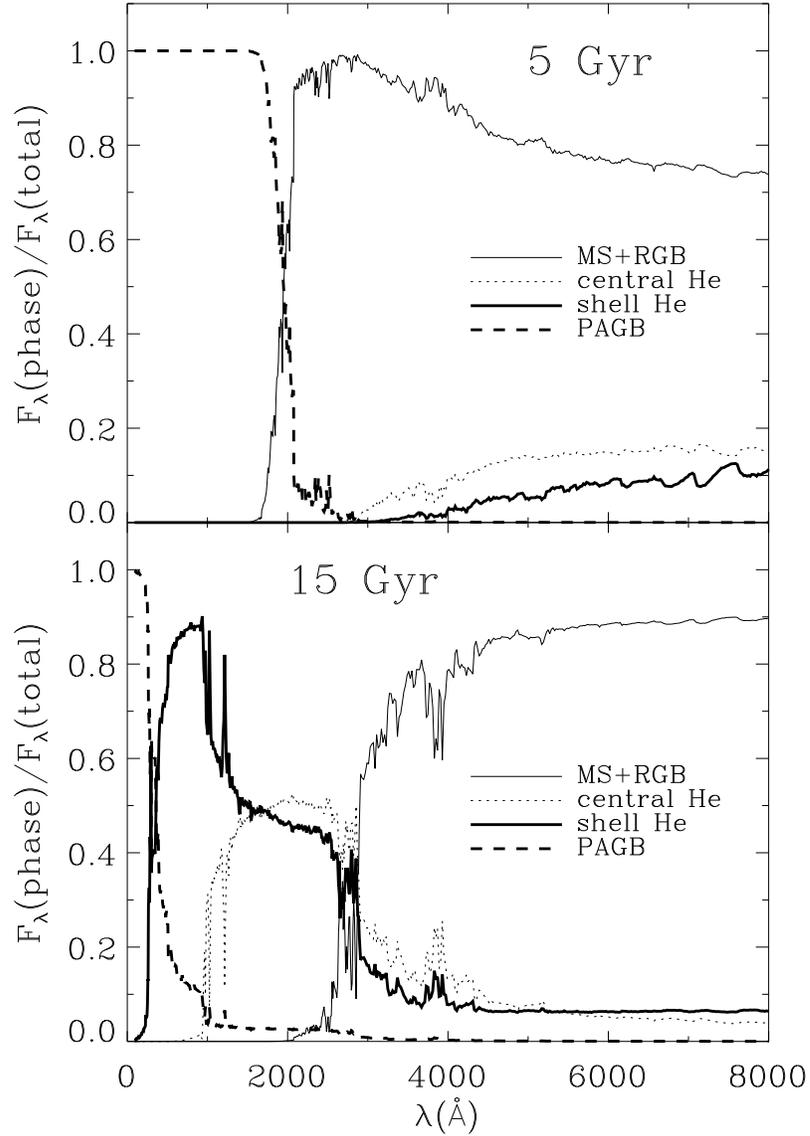}
\caption{
Light contribution from various evolutionary stages at different ages.
The upper and the lower panels are for 5 and 15~Gyrs old models, respectively,
both with ($Z$, $Y$, $\eta$, $x$, $\sigma$) = (0.02, 0.29, 1.0, 1.35, 0.06).
At an early age, the most UV light comes from PAGB stars, while the UV flux is
almost negligible, as seen in Figure 12. 
   The lower panel shows that much of the UV light at 15~Gyr, especially in
the far-UV, originates from shell helium-burning (evolved HB) stars in 
the UV-bright phase (SBP) discussed in YDK.
}\label{Figure 14}
\end{figure}

\clearpage

\begin{figure}
\plotone{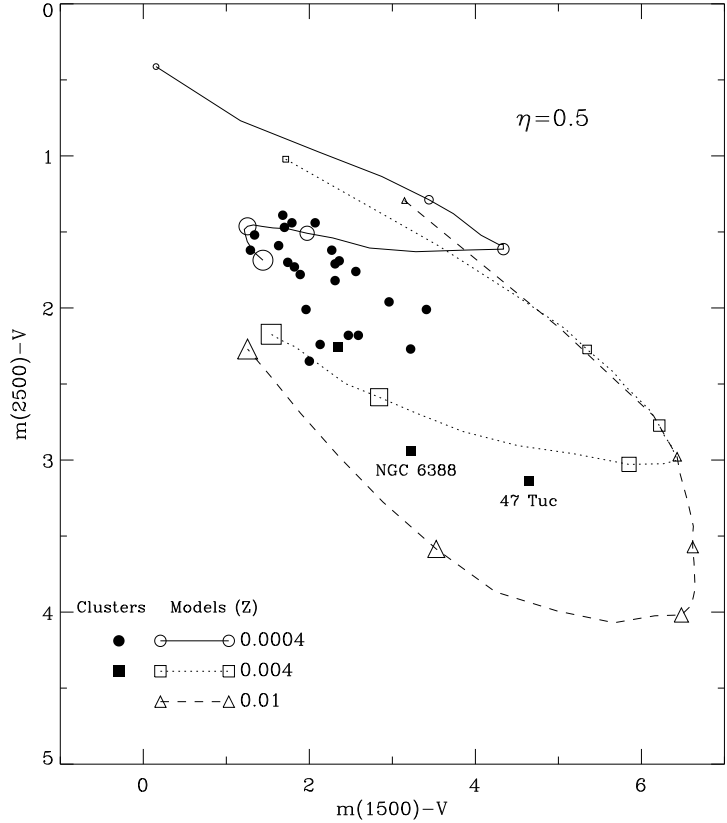}
\caption{
Effects of age and metallicity on the model two-color diagram, 
m(1500)-$V$ vs m(2500)-$V$. 
   Each line is for one metallicity but covering 1 -- 25~Gyrs of age. 
   From the smallest symbol to the largest, symbols denote 1, 5, 10, 15, 20, 
and 25~Gyr old models.
   In order to show only the effects of age and metallicity, other parameters 
are fixed, i.e. ($\eta$, $\sigma$, $x$) = (0.5, 0.06, 1.35). 
   The model colors are defined in the text, and the cluster data are from
Table 1 of Dorman et al. (1995). 
   Filled circles and filled squares are metal-poor cluster data ($Z <$ 0.002)
and metal-rich cluster data ($Z \geq 0.002$), respectively.
   Most data points are matched by the models with the age derived from
the MS turn-off in the CMD (approximately 15~Gyr).
   Because of the opacity effect, the more metal-rich, the redder the UV-to-$V$
colors. 
   The issue about the true $\eta$ is discussed in Section 4.2.
}\label{Figure 15}
\end{figure}

\clearpage

\begin{figure}
\plotone{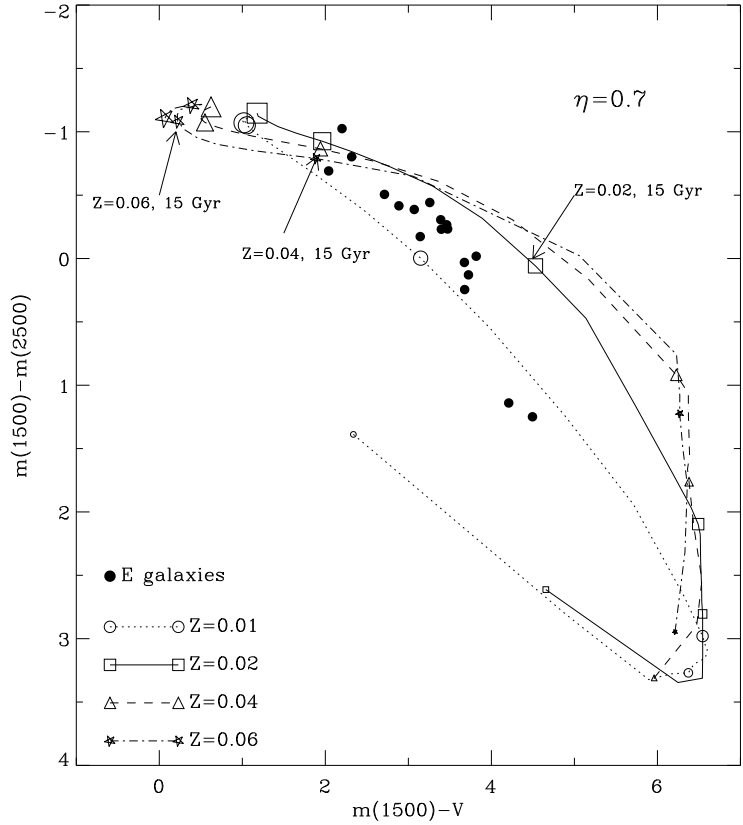}
\caption{
Same as Figure 15, but for metal-rich systems in the m(1500)-$V$ vs 
m(1500)-m(2500) diagram.
In order to show only the effects of age and metallicity, other parameters 
are fixed, i.e. ($\eta$, $\sigma$, $x$) = (0.7, 0.06, 1.35). 
The galaxy data are from Table 2 of Dorman et al. (1995).
If the majority of stars in elliptical galaxies are approximately 1 -- 2 \Zsun,
then the $\eta = 0.7$ models suggest that elliptical galaxies are about
15~Gyrs old.
}\label{Figure 16}
\end{figure}

\clearpage

\begin{figure}
\epsscale{0.7}
\plotone{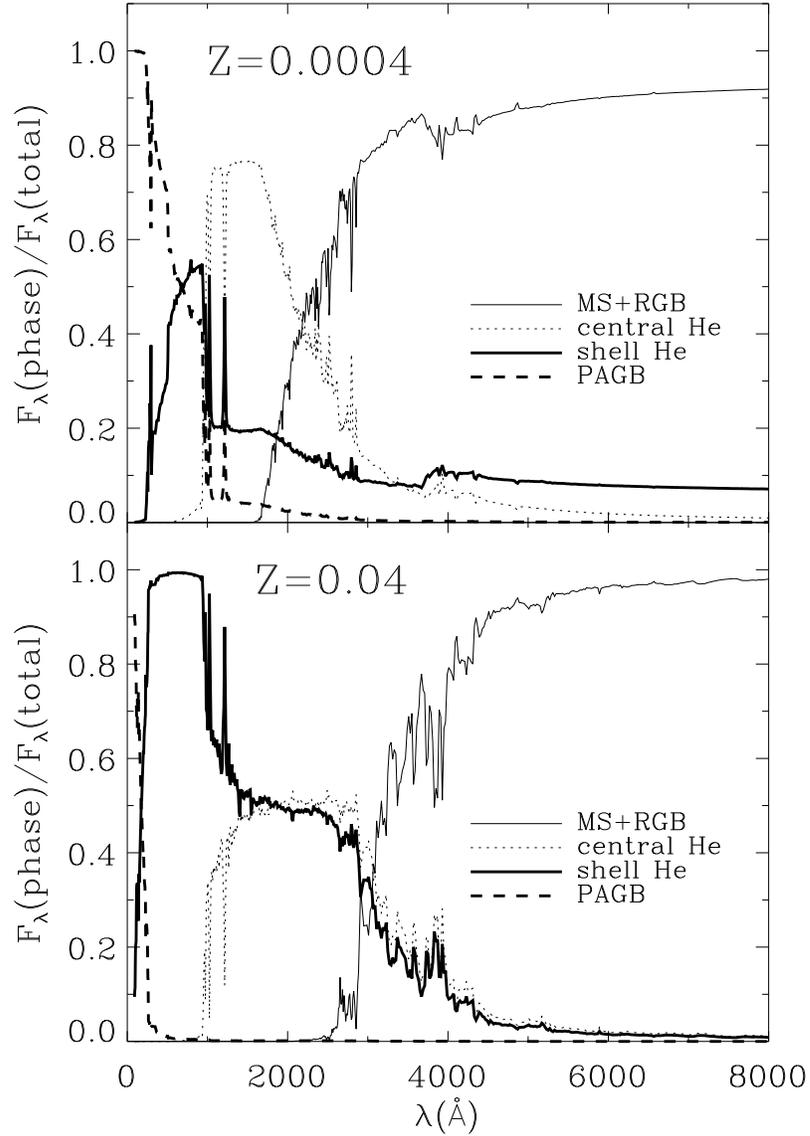}
\caption{
Light contribution from various evolutionary stages for different 
metallicities. The upper and lower panels are for a metal-poor model with
($Z$, $Y$) = (0.0004, 0.23), and a metal-rich model with (0.04, 0.35), 
respectively, both with (age, $\eta$, $x$, $\sigma$) = (15~Gyr, 1.0, 1.35, 
0.06).  Most UV light comes from the central helium-burning stars (HB stars) 
in the metal-poor model. But, a larger amount of UV light comes from the 
highly evolved, shell helium-burning stars in the metal-rich case 
(lower panel) due to the UV-bright phase that is more common in the 
metal-rich stars.
}\label{Figure 17}
\end{figure}

\clearpage

\begin{figure}
\plotone{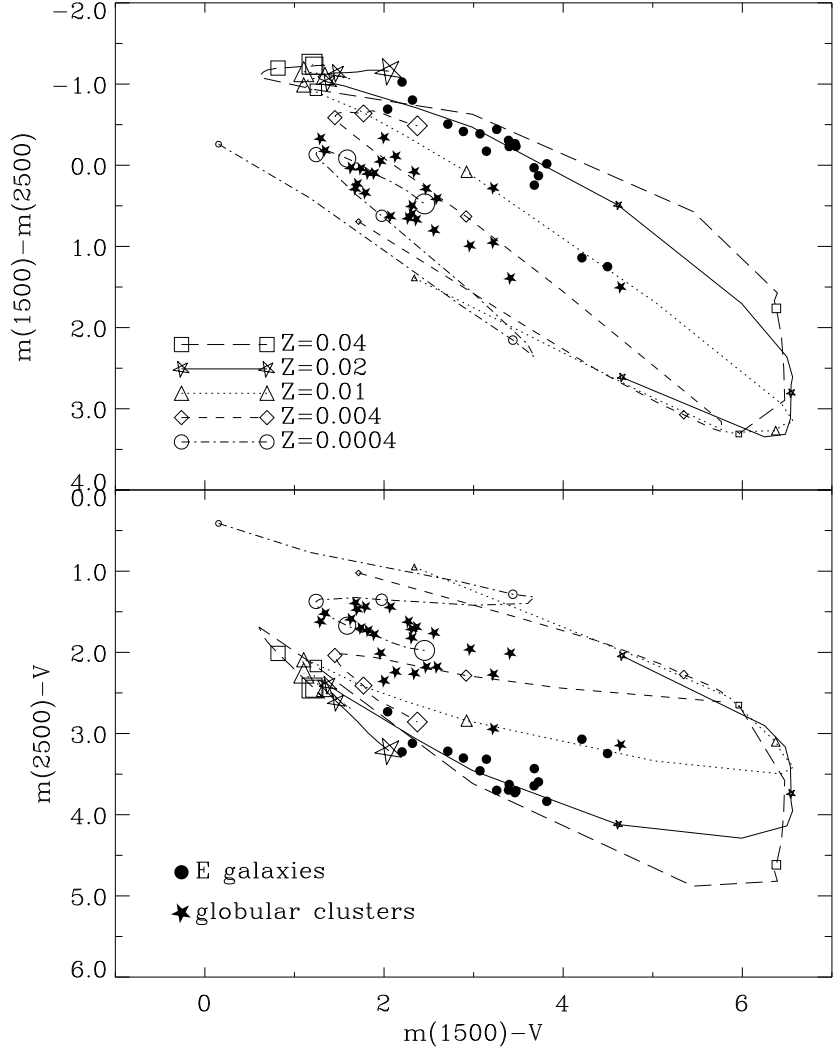}
\caption{
Effects of age and metallicity on model two-color diagrams. 
In order to show only the effects of age and metallicity, other parameters
are fixed as (\DYDZ, $\eta$, $\sigma$, $x$) = (3.0, 1.0, 0.06, 1.35).
If $\eta \approx$ 0.3 -- 0.5 for metal-poor stars as discussed in the text, 
the metal-poor models with $\eta = 1.0$ shown in this diagram are 
generating too much UV flux at a given age.
The issue about the true $\eta$ is discussed in Section 4.2.
Note that elliptical galaxies are best matched by
metal-rich models whereas metal-poor models reproduce the globular clusters
well as normally expected.
}\label{Figure 18}
\end{figure}

\clearpage

\begin{figure}
\epsscale{0.7}
\plotone{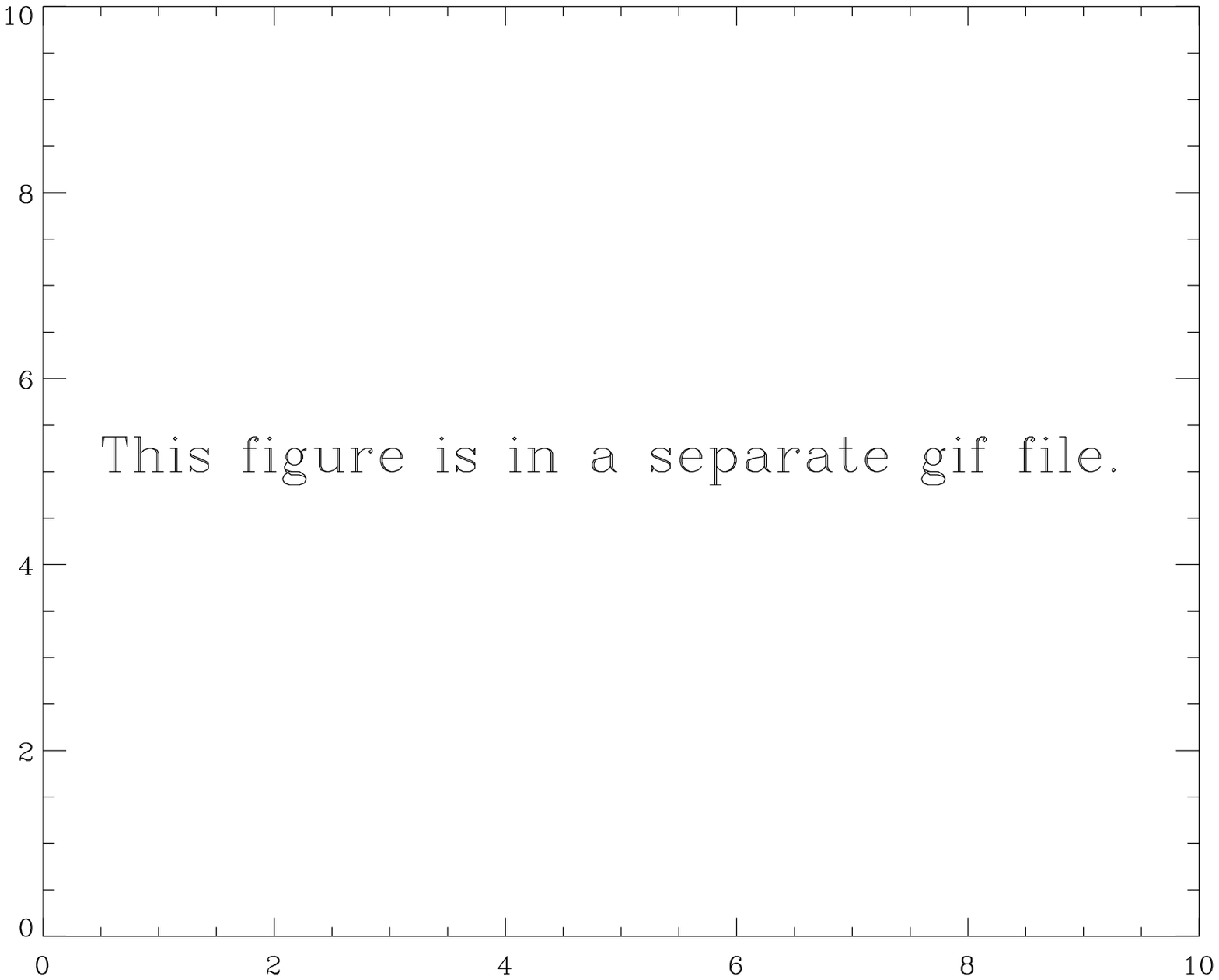}
\caption{
Effects of $\eta$ on model CMDs and SEDs for 
($Z$, $Y$, age, $\sigma$, $x$) = (0.04, 0.31, 15~Gyr, 0.06, 1.35).
$F_\lambda$ is in arbitrary but consistent unit.
It is clear that a higher mass loss efficiency leads to a larger number of
hot stars and a higher UV flux. Various studies favor a high $\eta$, namely, 
$\eta \geq 0.7$ for $Z \geq$ \Zsun. If this is true, old, metal-rich 
galaxies can exhibit a strong UV upturn within a Hubble time.
}\label{Figure 19}
\end{figure}

\clearpage

\begin{figure}
\epsscale{0.7}
\plotone{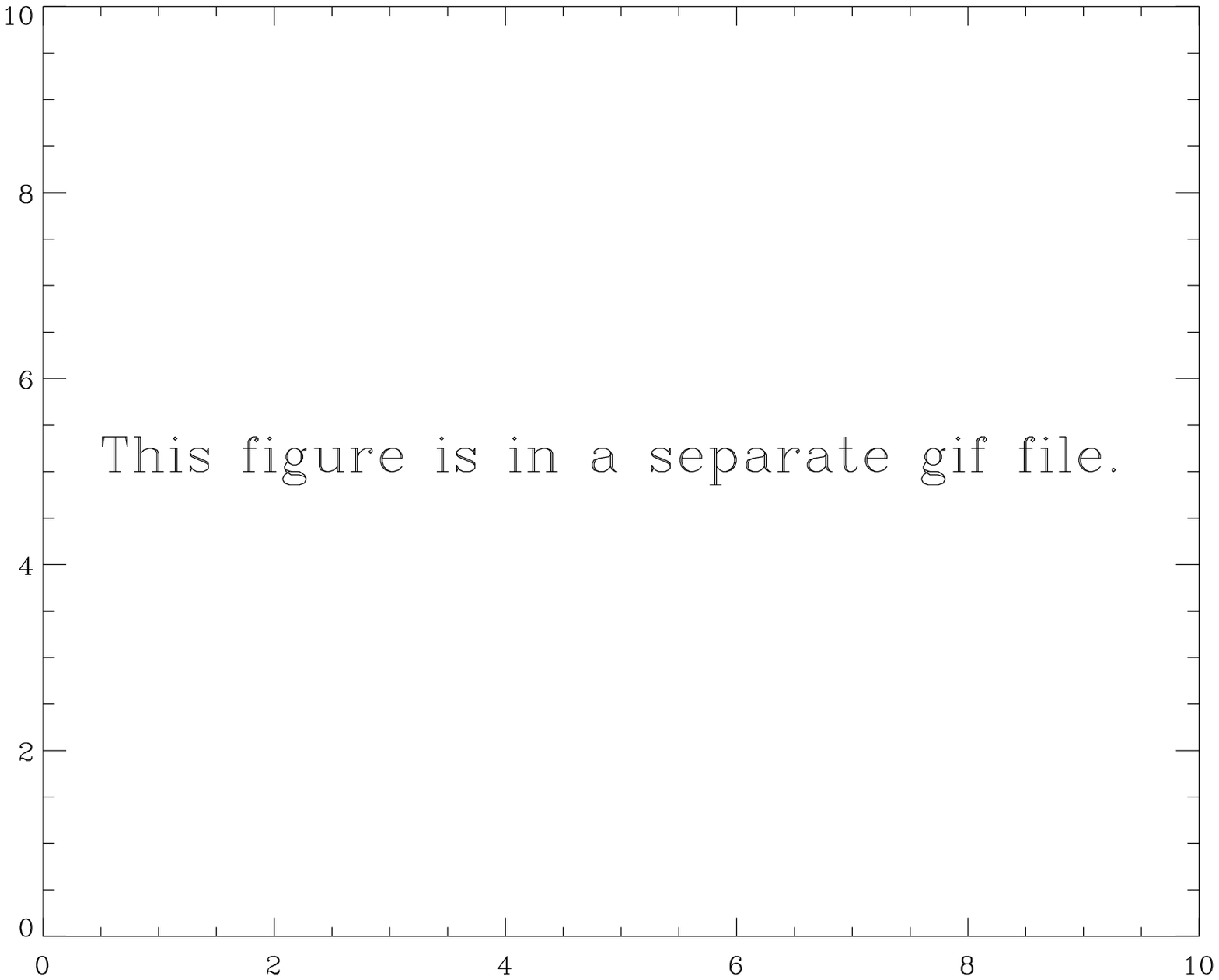}
\caption{
Effects of $\eta$ on model CMDs and SEDs for 
($Z$, $Y$, age, $\sigma$, $x$) = (0.0004, 0.23, 15~Gyr, 0.06, 1.35).
$F_\lambda$ is in arbitrary but consistent unit.
Unless $\eta$ were as high as unity, an
unlikely high value for metal-poor stars, metal-poor populations would
fail to produce a strong UV upturn.
}\label{Figure 20}
\end{figure}

\clearpage

\begin{figure}
\plotone{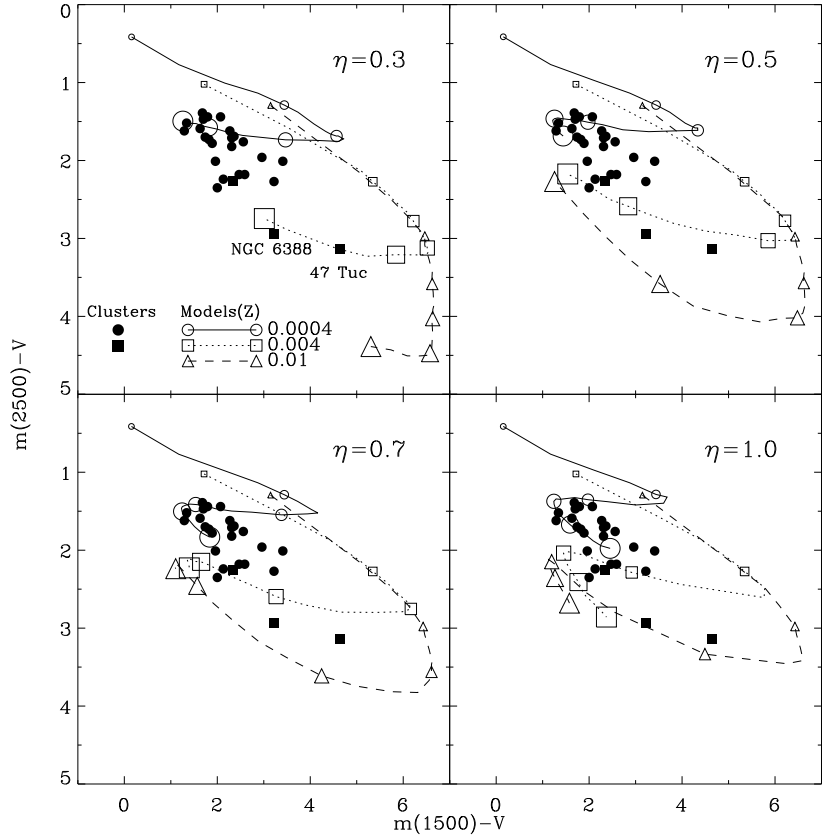}
\caption{
Effects of $\eta$ on model two-color diagrams, m(1500)-$V$ vs m(2500)-$V$. 
See the Figure 15 caption for model and symbol descriptions.
In order to show only the effects of $\eta$, other parameters are fixed, i.e.
($\sigma$, $x$) = (0.06, 1.35). Data for metal-poor globular clusters 
(filled circles) and relatively metal-rich globular clusters (filled squares) 
are shown. Note that $\eta = 0.3$ models, which reasonably reproduce the
colors of metal-poor clusters, fail to match the colors of the metal-rich 
globular clusters (e.g. NGC\,6388, $Z \approx$ 0.006) at an acceptable age; 
the $\eta = 0.3$ model suggests an age of about 24~Gyrs for NGC\,6388.
}\label{Figure 21}
\end{figure}

\clearpage

\begin{figure}
\plotone{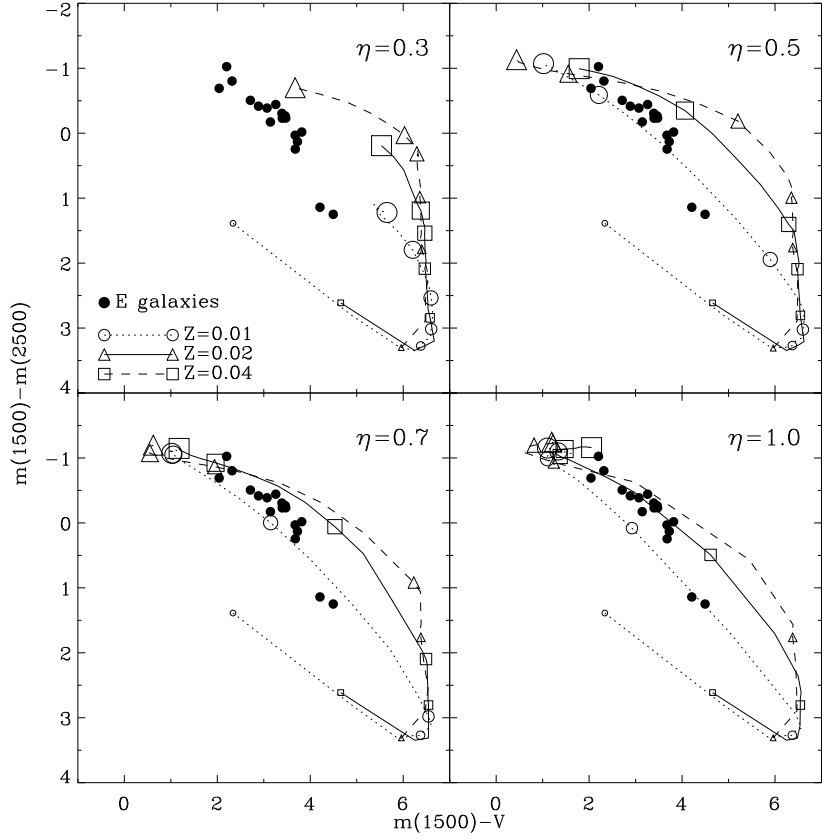}
\caption{
Same as Figure 21, but for metal-rich systems on the m(1500)-$V$ vs 
m(1500)-m(2500) plane.
The UV strength of elliptical galaxies cannot be reproduced by low
$\eta$ models. If elliptical galaxies are approximately 15~Gyrs old, this
suggests $\eta \gtrsim 0.7$ for metal-rich stars.
}\label{Figure 22}
\end{figure}

\clearpage

\begin{figure}
\plotone{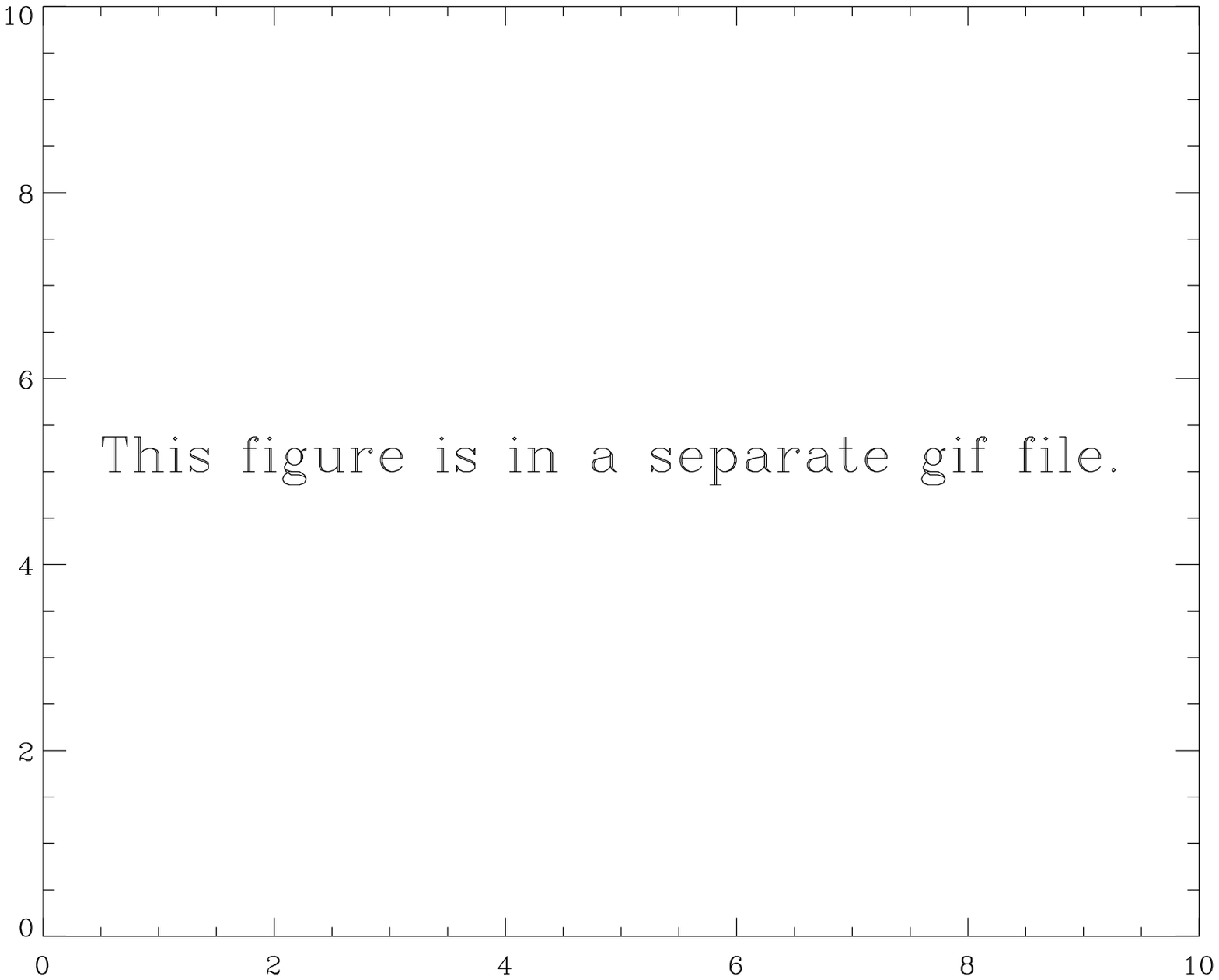}
\caption{
An example showing the effects of \DYDZ on model CMDs and SEDs
for ($Z$, age, $\eta$, $\sigma$, $x$) = (0.02, 13~Gyr, 0.06, 1.0, 1.35).
$F_\lambda$ is in arbitrary but consistent unit.
A model with a higher \DYDZ shows a higher UV flux because (1) HB stars with
a higher helium abundance, $Y$, become UV-bright more easily, and (2) stars 
with a higher $Y$ evolve faster on MS and RGB when $Z \gtrsim$ \Zsun. 
Thus, a galaxy with a higher \DYDZ contains a larger number of 
low-mass (hot) HB stars at a fixed age. 
}\label{Figure 23}
\end{figure}
    
\clearpage

\begin{figure}
\plotone{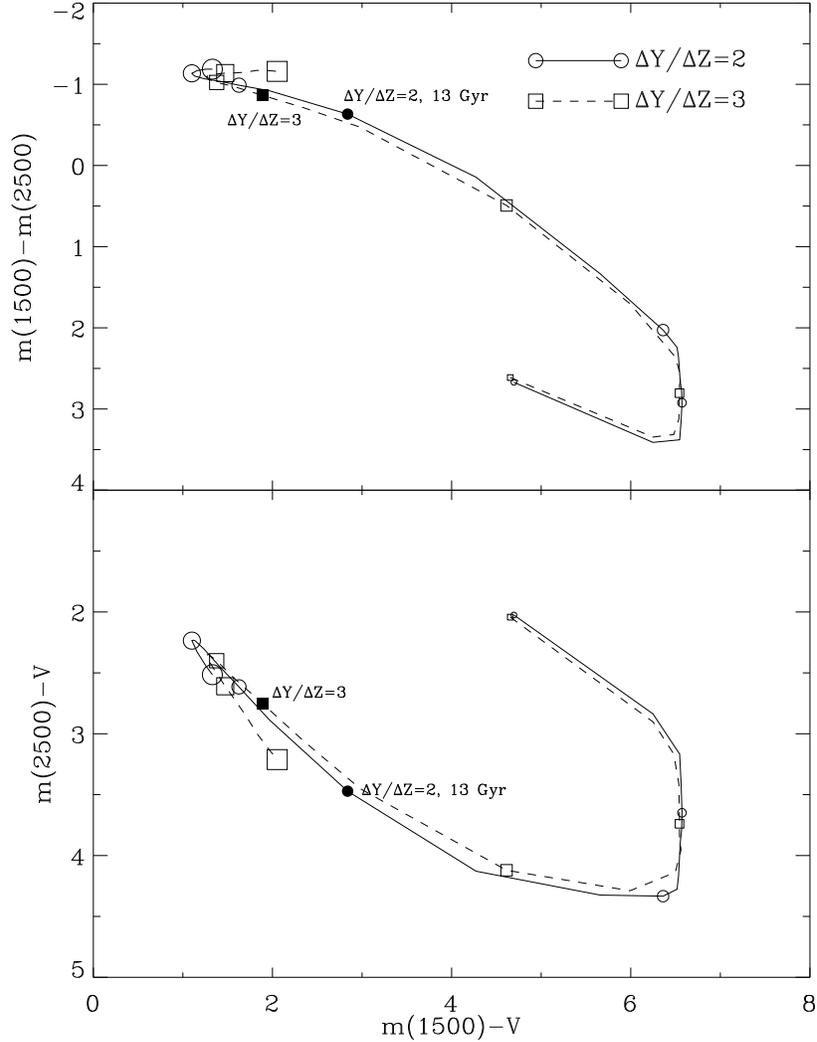}
\caption{
Effects of \DYDZ on model two-color diagrams. 
Models are for ($Z$, $\eta$, $\sigma$, $x$) = (0.02, 1.0, 0.06, 1.35). 
Higher-\DYDZ models generate UV flux more quickly because stars with 
higher helium abundance evolve faster and because the UV-bright phase of 
metal-rich HB stars is more significant as helium abundance increases 
(see YDK). But it is difficult to choose the true \DYDZ by UV color fitting  
unless the age is known a priori, because, if \DYDZ is within
the range of 2 -- 3, the model sequences of different values of \DYDZ look 
very much alike. Symbols are the same as in the Figure 15 except that
13~Gyr-old models (filled symbols) are also marked to be compared with
Figure 23.
}\label{Figure 24}
\end{figure}

\clearpage

\begin{figure}
\plotone{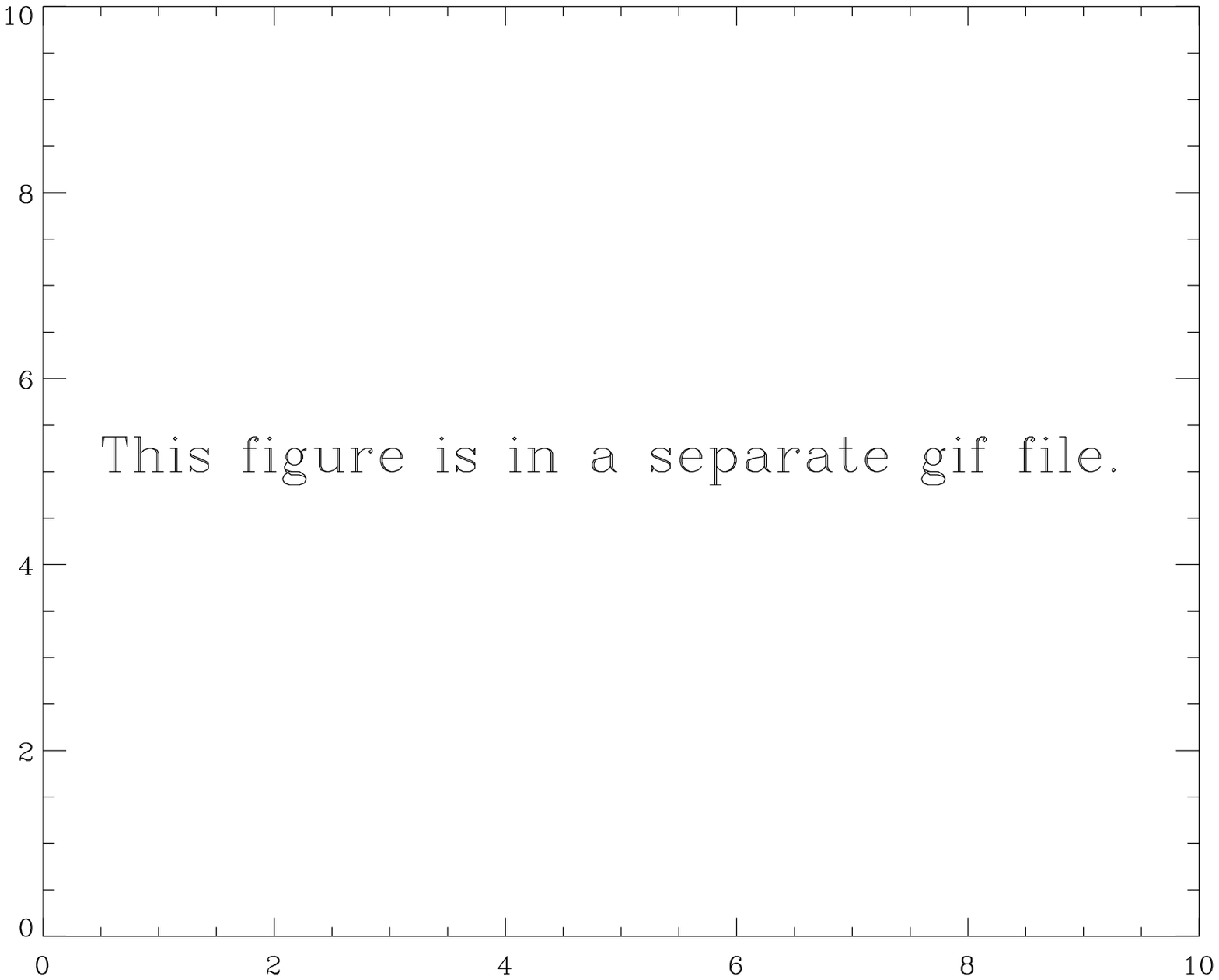}
\caption{
Effects of the IMF slope $x$ on model CMDs and SEDs for 
($Z$, $Y$, age, $\sigma$, $\eta$) = (0.02, 0.29, 15~Gyr, 0.06, 1.0).
$F_\lambda$ is in arbitrary but consistent unit.
A model SED is not very sensitive to the IMF slope $x$.
}\label{Figure 25}
\end{figure}

\clearpage

\begin{figure}
\plotone{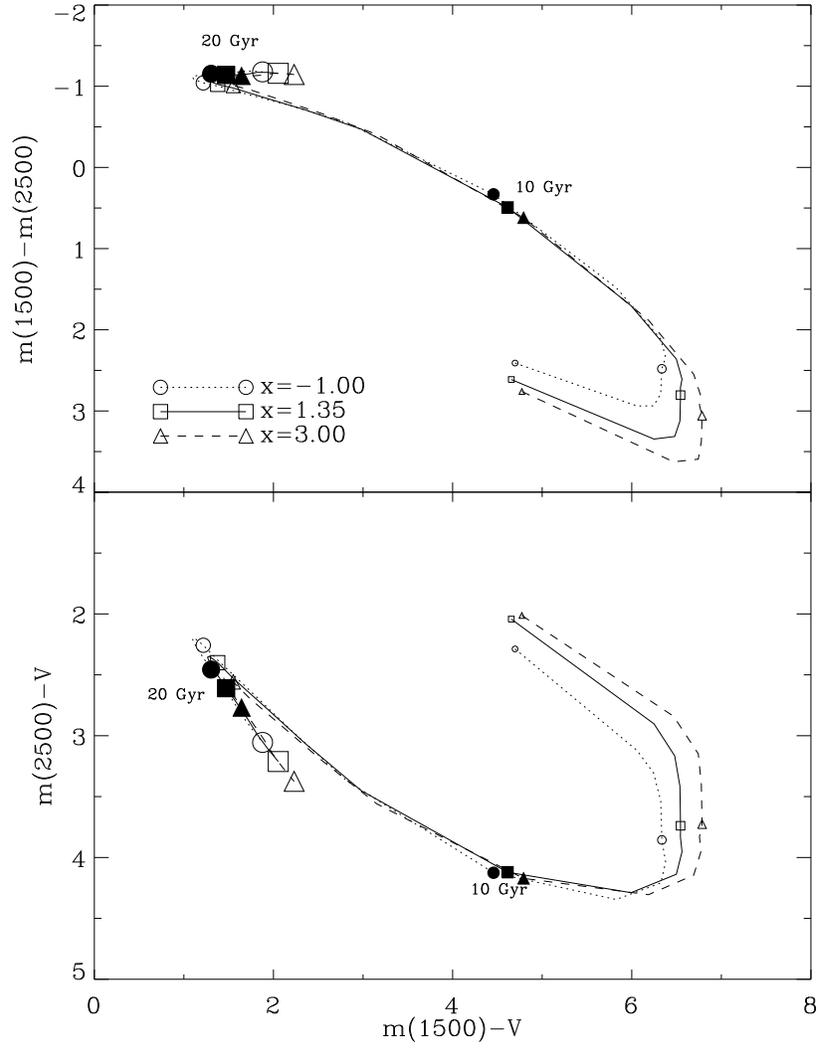}
\caption{
Effects of the IMF slope $x$ on model two-color diagrams. 
Models are for ($Z$, $Y$, $\eta$, $\sigma$) = (0.02, 0.29, 1.0, 0.06). 
Note that the IMF slope has little effect on the UV-to-$V$ colors.
See the Figure 15 caption for details about symbols.
}\label{Figure 26}
\end{figure}

\clearpage

\begin{figure}
\plotone{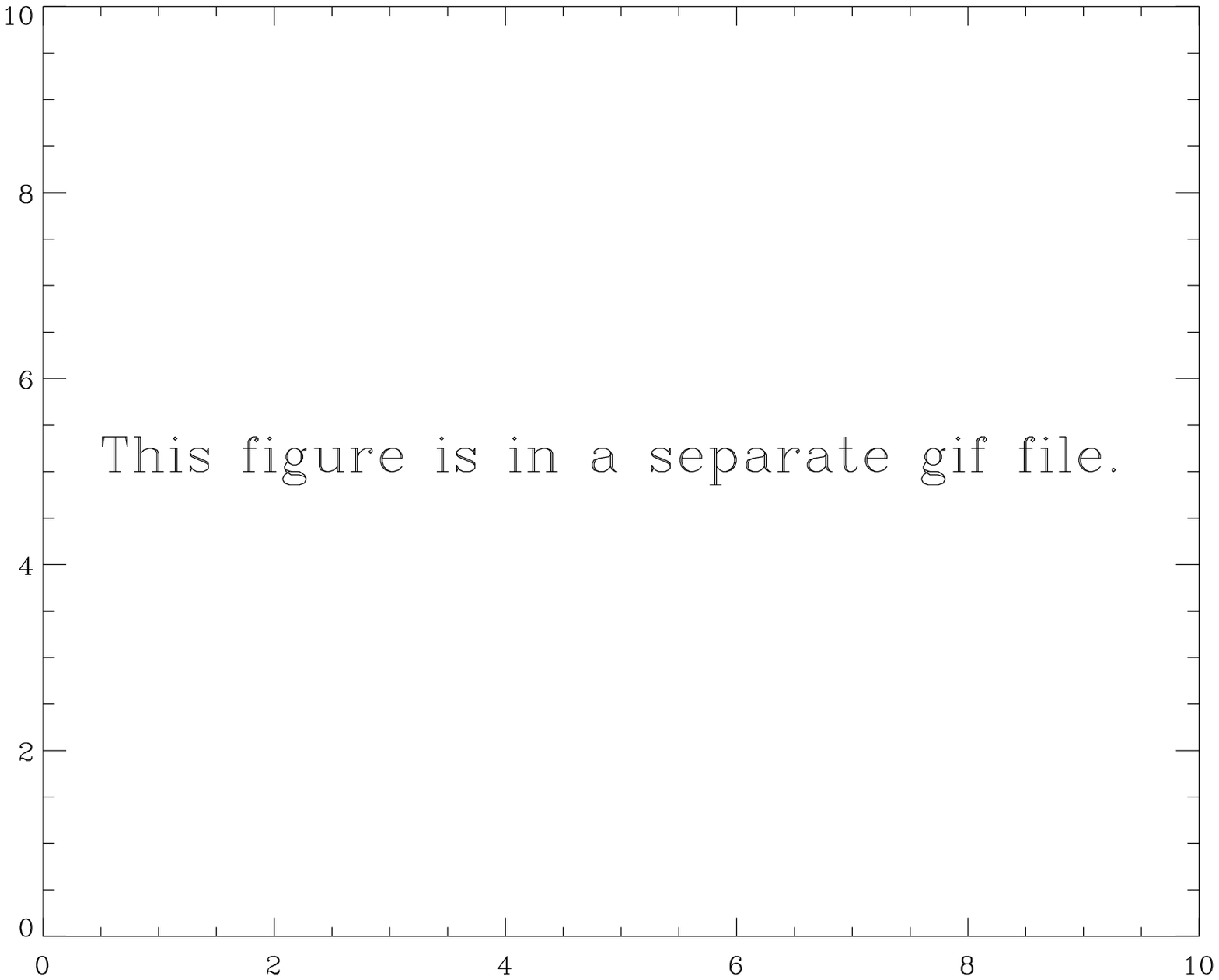}
\caption{
Effects of the gaussian mass dispersion factor $\sigma$ on model 
CMDs and SEDs for ($Z$, $Y$, age, $\sigma$, $x$) = (0.02, 0.29, 10~Gyr, 
0.06, 1.35). $F_\lambda$ is in arbitrary but consistent unit.
When a galaxy is so young that a larger fraction of its 
core helium-burning stars is cool, a larger $\sigma$ causes a higher UV flux.
}\label{Figure 27}
\end{figure}

\clearpage

\begin{figure}
\plotone{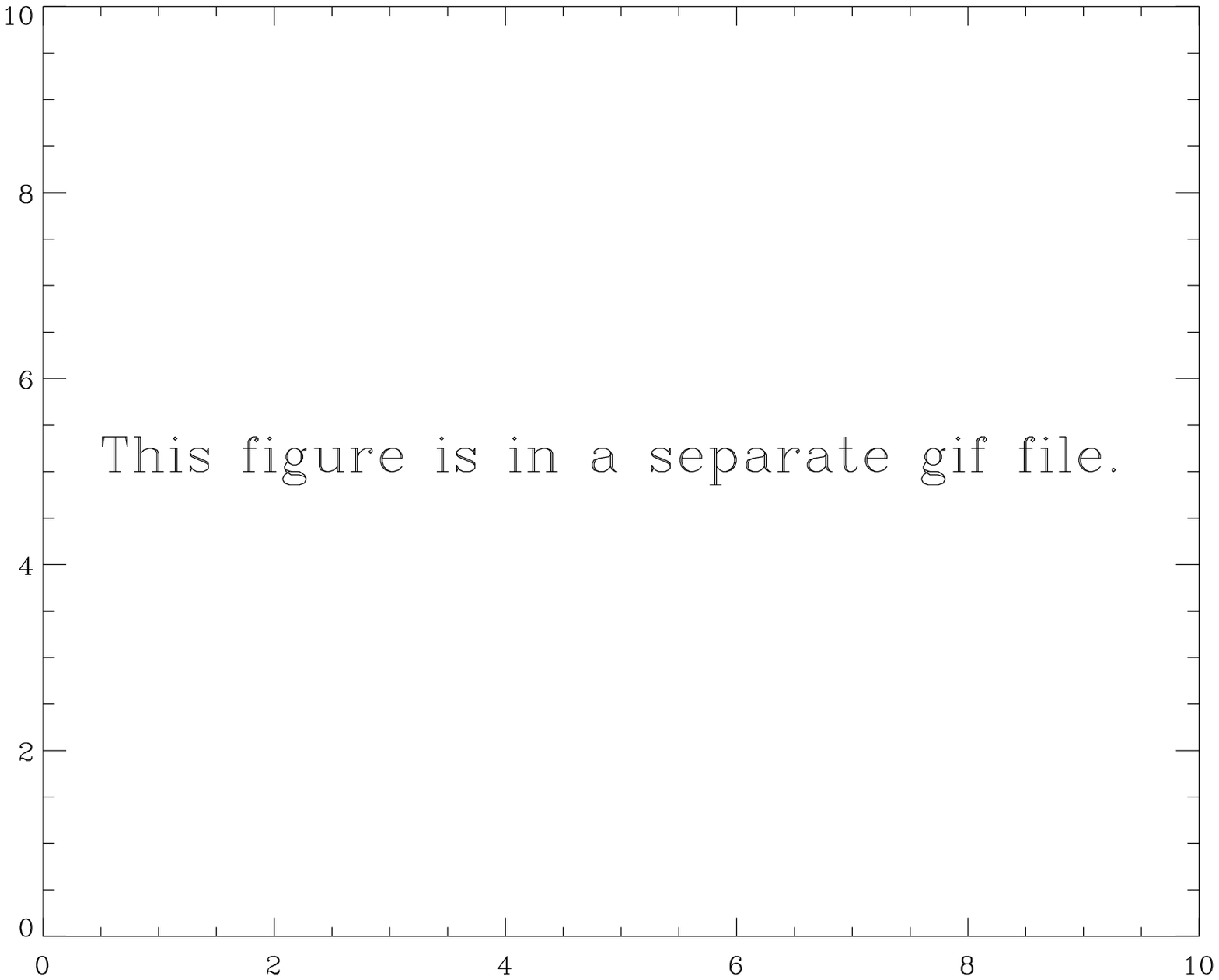}
\caption{
Same as Figure 27, but for a 15~Gyr model galaxy. When a galaxy is old 
enough that a larger fraction of its core helium-burning stars is hot,
a larger $\sigma$ causes a lower UV flux.
}\label{Figure 28}
\end{figure}

\clearpage

\begin{figure}
\plotone{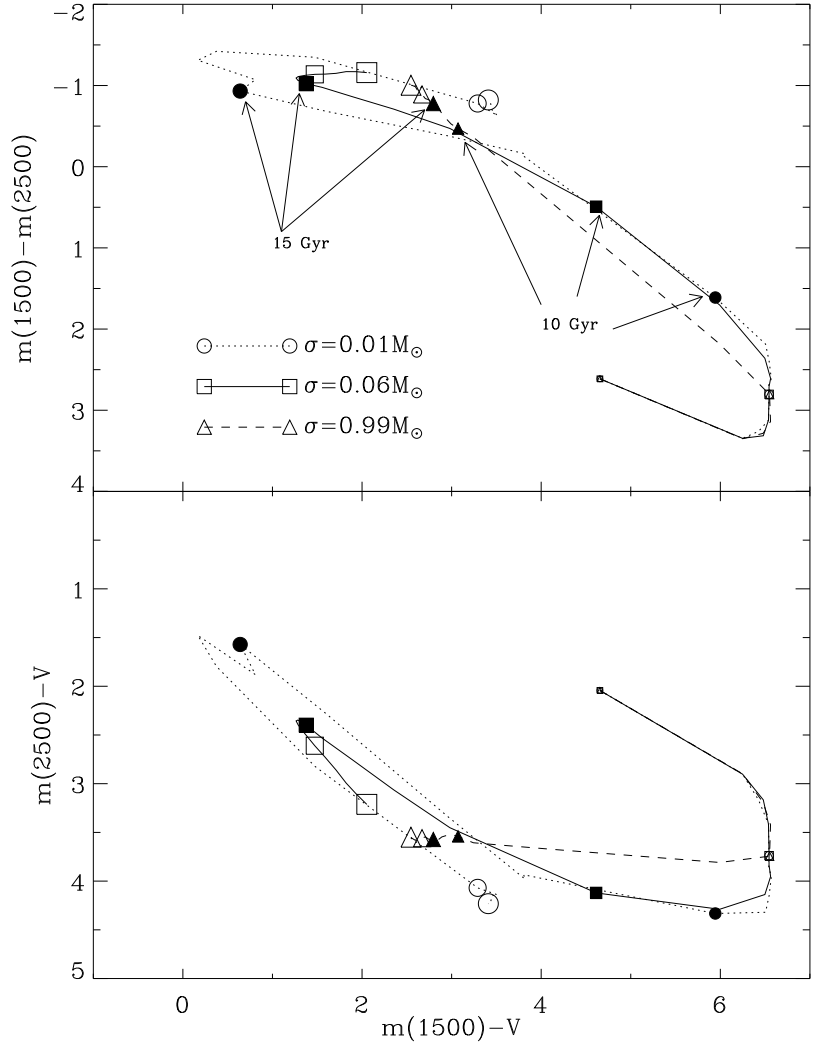}
\caption{
Effects of the mass dispersion parameter, $\sigma$, on model two-color 
diagrams. 
All models are for ($Z$, $Y$, $\eta$, $x$) = (0.02, 0.29, 1.0, 1.35). 
It is clear that the model UV-to-$V$ colors are sensitive to $\sigma$ which 
determines the width of the HB.
Thus, oversimplistic HB treatments are likely to lead to wrong conclusions.
See the Figure 15 caption for details about symbols.
}\label{Figure 29}
\end{figure}


\clearpage

\begin{thebibliography}{}
\bibitem[Aaronson \& Mould 1982]{am82}  Aaronson, M., \& Mould, J. 1982, \apjsupp, 48, 161
\bibitem[Allen 1976]{a76}  Allen, C. W. 1976, in Astrophysical Quantities (The Athlone Press: London), 202
\bibitem[Bailyn 1995]{bai95} Bailyn, C. D. 1995, Ann. Rev. A. \& Ap, 33, 133
\bibitem[Bertola et al. 1993]{ber93} Bertola, F., Burstein, D., Buson, L. M., \& Renzini, A. 1993, \apj, 403, 577
\bibitem[Bl\"{o}cker \& Sch\"{o}nberner 1990]{bs90} Bl\"{o}cker, T., \& Sch\"{o}nberner, D. 1990, \aap, 240, L11
\bibitem[Bowen \& Willson 1991]{bw91} Bowen, G. H., \& Willson, L. A. 1991, \apj, 371, L53
\bibitem[Bressan et al. 1994]{bcf94}  Bressan, A., Chiosi, C., \& Fagotto, F. 1994, \apjsupp, 94, 63
\bibitem[Brown et al. 1995]{bfd95}  Brown, T. H., Ferguson, H. C., \& Davidsen, A. F. 1995, \apj, 454, L15
\bibitem[Brown et al. 1997]{bfdd97}  Brown, T. M., Ferguson, H. C., Davidsen, A. F., \& Dorman, B. 1997, preprint
\bibitem[Bruzual \& Charlot 1993]{bc93} Bruzual, A. G., \& Charlot, S. 1993, \apj, 405, 538
\bibitem[Burstein et al. 1988]{b88}  Burstein, D., Bertola, F., Buson, L. M., Faber, S. M., \& Lauer, T. R. 1988, \apj, 328, 440
\bibitem[Buser \& Kurucz 1992]{bk92}  Buser, R., \& Kurucz, R. L. 1992, \aap, 264, 557
\bibitem[Buzzoni 1995]{b95}  Buzzoni, A. 1995, \apjsupp, 98, 69
\bibitem[Cacciari et al. 1995]{cac95}  Cacciari, C., Fusi Pecci, F., Bragaglia, A., \& Buzzoni, A. 1991, \aap, 301, 684
\bibitem[Castellani \& Castellani 1993]{cc93} Castellani, M., \& Castellani, V. 1993, \apj, 407, 649
\bibitem[Castellani \& Tornamb\'{e} 1991]{ct91}  Castellani, M., \& Tornamb\'{e}, A. 1991, \apj, 381, 393
\bibitem[Catelan 1993]{c93}  Catelan, M. 1993, \aap Suppl., 98, 547
\bibitem[Chaboyer et al. 1996]{cds96}  Chaboyer, B., Demarque, P., \& Sarajedini, A. 1996, \apj, 459, 558
\bibitem[Charlot et al. 1996]{cwb96}  Charlot, S., Worthey, G., \& Bressan, A. 1996, \apj, 457, 625
\bibitem[Clement et al. 1991]{cks91} Clement, C. M., Kinman, T. D., \& Suntzeff, N. B. 1991, \apj, 372, 273
\bibitem[Code \& Welch 1979]{cw79} Code, A. D., \& Welch, G. A. 1979, \apj, 228, 95
\bibitem[Cole \& Deupree 1980]{cd80}  Cole, P. W., \& Deupree, R. G. 1980, \apj, 239, 284
\bibitem[Cole et al. 1985]{cdd85}  Cole, P. W., Demarque, P. \& Deupree, R. G. 1985, \apj, 291, 291
\bibitem[Cox 1991]{c91}  Cox, A. N. 1991, \apj, 381, L71
\bibitem[de Boer et al. 1997]{dts97} de Boer, K. S., Tucholke, H.-J., \& Schmidt, J. H. K. 1997, \aap, 317, L23
\bibitem[Demarque et al. 1996]{d96} Demarque, P., Chaboyer, B., Guenther, D., Pinsonneault, L., Pinsonneault, M., \& Yi, S. 1996, The Yale Isochrones 1996 
\bibitem[Demarque \& Mengel 1971]{dm71} Demarque, P. \& Mengel, J. G. 1971 \apj, 164, 469
\bibitem[Demarque \& Pinsonneault 1988]{dp88} Demarque, P., \& Pinsonneault, M. H. 1988, in Progress \& Opportunities in Southern Hemisphere Optical Astronomy, eds. V. M. Blanco \& M. M. Phillips (San Francisco: ASP), 371
\bibitem[Dixon et al. 1996]{dddf96}  Dixon, W. V. D., Davidsen, A. F., Dorman, B., \& Ferguson, H. 1996, \aj, in press
\bibitem[Dorman et al. 1993]{dro93}  Dorman, B., Rood, R. T., \& O'Connell, R. 1993, \apj, 419, 596
\bibitem[Dorman et al. 1995]{dor95}  Dorman, B., O'Connell, R., \& Rood, R. T. 1995, \apj, 442, 105
\bibitem[Dupree 1986]{d86}  Dupree, A. K. 1986, \araa, 24, 377
\bibitem[D'Cruz et al. 1996]{ddro96}  D'Cruz, N. L., Dorman, B., Rood, R. T., \& O'Connell, R. 1996, \apj, 466, 359
\bibitem[Faber 1972]{f72}  Faber. S. M. 1972, \aap, 20, 361
\bibitem[Faber 1983]{f83}  Faber. S. M. 1983, Highlights Astr., 6, 165
\bibitem[Fagotto et al. 1994]{fag94}  Fagotto, F., Bressan, A., Bertelli, G., \& Chiosi, C. 1994, A\&AS, 105, 39
\bibitem[Fanelli et al. 1987]{fot87}  Fanelli, M. N., O'Connell, R. W., \& Thuan, T. X. 1987, \apj, 321, 768
\bibitem[Fanelli et al. 1992]{fan92}  Fanelli, M. N., O'Connell, R. W., Burstein, D., \& Wu, C.-C. 1992, \apjsupp, 82, 197
\bibitem[Ferguson \& Davidsen 1993]{fd93}  Ferguson, H. C., \& Davidsen, A. F. 1993, \apj, 408, 92
\bibitem[Ferraro et al. 1992]{fer92}  Ferraro, F. R., Clementi, G., Fusi Pecci, F., Sortino, R., \& Buonanno, R. 1992, \mnras, 256, 391
\bibitem[Fusi Pecci et al. 1993]{fusi93}  Fusi Pecci, F., Ferraro, F. R., Bellazzini, M., Djorgovski, S., Piotto, G., \& Buonanno, R. 1993, \aj, 105, 1145
\bibitem[Greggio \& Renzini 1990]{gr90}  Greggio, L., \& Renzini, A. 1990, \apj, 364, 35
\bibitem[Gunn \& Stryker 1983]{gs83}  Gunn, J. E., \& Stryker, L. L. 1983, \apjsupp, 52, 121
\bibitem[Gunn et al. 1981]{gst81}  Gunn, J. E., Stryker, L. L., \& Tinsley, B. M. 1981, \apj, 249, 48
\bibitem[Gustafsson \& J\mbox{\o}rgensen 1994]{gj94}  Gustafsson, B., \& J\mbox{\o}rgensen, U. G. 1994, Astron. Astrophys. Rev., 6, 19
\bibitem[Hayashi et al. 1962]{h62}  Hayashi, C., Hoshi, R., \& Sugimoto, D. 1962, Prog. Theor. Phys. Suppl., 22, 1
\bibitem[Horch et al. 1992]{hdp92}  Horch, E., Demarque, P., \& Pinsonneault, M. 1992, \apj, 388, L53
\bibitem[Hubeny 1988]{h88}  Hubeny, I. 1988, Computer Physics Communications, 52, 103
\bibitem[J\mbox{\o}rgensen \& Thejll 1993a]{jt93a} J\mbox{\o}rgensen, U. G., \& Thejll, P. 1993, \aap, 272, 255
\bibitem[J\mbox{\o}rgensen \& Thejll 1993b]{jt93b} J\mbox{\o}rgensen, U. G., \& Thejll, P. 1993, \apj, 411, L67
\bibitem[Kaluzny \& Udalski 1992]{ku92}  Kaluzny, J., \& Udalski, A. 1992, AcA, 42, 29
\bibitem[Kaluzny \& Rucinski 1995]{kr95}  Kaluzny, J., \& Rucinski, S. M. 1995, \aap, 114, 1
\bibitem[Kirkpatrick et al. 1993]{k93}  Kirkpatrick, J. D., Kelly, D. M., Rieke, G. H., Liebert, J., Allard, F., \& Wehrse, R. 1993, \apj, 402, 643
\bibitem[Kovacs et al. 1991]{kbm91}  Kovacs, G., Buchler, J. R., \& Marom, A. 1991, \aap, 252, 27
\bibitem[Kudritzki \& Reimers 1978]{kr78}  Kudritzki, R. P., \& Reimers, D. 1978, \aap, 70, 227
\bibitem[Kurucz 1992]{k92}  Kurucz, R. 1992, in The Stellar Population in Galaxies, ed. B. Barbuy \& A. Renzini (Dordrecht: Reidel), 225
\bibitem[Larson 1992]{l92}  Larson, R. B. 1992, \mnras, 256, 641
\bibitem[Larson 1995]{l95}  Larson, R. B. 1995, \mnras, 272, 213
\bibitem[Lee \& Demarque 1990]{ld90}  Lee, Y.-W., \& Demarque, P. 1990, \apjsupp, 73, 709
\bibitem[Lee et al. 1990]{ldz90}  Lee, Y.-W., Demarque, P., \& Zinn, R. 1990, \apj, 350, 155
\bibitem[Lee et al. 1994]{ldz94}  Lee, Y.-W., Demarque, P., \& Zinn, R. 1994, \apj, 423, 248
\bibitem[Liebert et al. 1994]{lsg94} Liebert, J., Saffer, R. A., \& Green, E. M. 1994, \aj, 107, 1408
\bibitem[Maeder 1991]{m91} Maeder, A. 1991, QJRAS, 32, 217
\bibitem[Magris \& Bruzual 1993]{mb93} Magris, G. C., \& Bruzual, A. G. 1993, \apj, 417, 102
\bibitem[McClure \& van den Bergh 1968]{mv68}McClure, R. D., \& van den Bergh, S. 1968, \aj, 73, 313
\bibitem[Miller \& Scalo 1979]{ms79} Miller, G. E., \& Scalo, J. M. 1979, \apjsupp, 41, 513
\bibitem[Moehler et al. 1995]{mhd95} Moehler, S., Heber, U., \& de Boer, K. S. 1995, \aap, 294, 65
\bibitem[Morossi et al. 1993]{m93} Morossi, C., Franchini, M., Malagnini, M. L., Kurucz, R. L., \& Buser, R. 1993, \aap, 277, 173
\bibitem[Mould \& Aaronson 1982]{ma82}  Mould, J., \& Aaronson, M. 1982, \apj, 263, 629
\bibitem[Nesci \& Perola 1985]{np85} Nesci, R., \& Perola, G. C. 1985, \aap, 145, 296
\bibitem[O'Connell 1976]{o76}  O'Connell, R. W. 1976, \apj, 206, 370
\bibitem[O'Connell 1987]{o87}  O'Connell, R. W. 1987, in Stellar Populations, ed. C. A. Norman, A. Renzini, \& M. Tosi (Cambridge: Cambridge University Press), 167
\bibitem[O'Connell et al. 1992]{o92}  O'Connell, R. W. et al. 1992, \apj, 395, L45
\bibitem[Park \& Lee 1997]{pl97}  Park, J.-H., \& Lee, Y.-W. 1997, \apj, 476, in press
\bibitem[Petersen 1973]{p73}  Petersen, J. O. 1973, \aap, 27, 89
\bibitem[Pickles 1985a]{p85a}  Pickles, A. J. 1985a, \apjsupp, 59, 33
\bibitem[Pickles 1985b]{p85b}  Pickles, A. J. 1985b, \apj, 296, 340
\bibitem[Refsdal \& Weigert 1970]{rw70}  Refsdal, S., \& Weigert, A. 1970, \aap, 6, 426
\bibitem[Reimers 1975]{r75}  Reimers, D. 1975, M\'{e}m. Soc. Roy. Sci. Li\`{e}ge, 6th Ser., 8, 369
\bibitem[Renzini 1981]{r81}  Renzini, A. 1981, in Physical Processes in Red Giants, ed. I. Iben \& A. Renzini (Dordrecht: Reidel), 431
\bibitem[Rood 1973]{r73}  Rood, R. T. 1973, \apj, 184, 815
\bibitem[Rood 1990]{r90}  Rood, R. T. 1990, in The Confrontation between Stellar Pulsation and Evolution, ed. C. Cacciari \& G. Clementi (San Francisco: ASP), 11
\bibitem[Salpeter 1955]{s55} Salpeter E. E. 1955, \apj, 121, 161
\bibitem[Sch\"{o}nberner 1979]{s79}  Sch\"{o}nberner, D. 1979, \aap, 79, 108
\bibitem[Sch\"{o}nberner 1983]{s83}  Sch\"{o}nberner, D. 1983, \apj, 272, 708
\bibitem[Silva \& Cornell 1992]{sc92}  Silva, D. R., \& Cornell, M. E. 1992, \apjsupp, 81, 865
\bibitem[Smith \& Norris 1983]{sn83} Smith, G., \& Norris, J. 1983, \apj, 264, 215
\bibitem[Spinrad \& Taylor 1971]{st71} Spinrad, H., \& Taylor, B. 1971, \apjsupp, 22, 445
\bibitem[Sweigart 1987]{s87}  Sweigart, A. V. 1987, \apjsupp, 65, 95
\bibitem[Sweigart et al. 1974]{smd74}  Sweigart, A. V., Mengel, J. G., \& Demarque, P. 1974, \aap, 30, 13
\bibitem[Tantalo et al. 1996]{tan96} Tantalo, R., Chiosi, C., Bressan, A., \& Fagotto, F. 1996, \aap, 311,361
\bibitem[Tinsley 1968]{t68} Tinsley, B. M. 1968, \apj, 151, 547
\bibitem[Tinsley 1972]{t72} Tinsley, B. M. 1972, \aap, 20, 383
\bibitem[Tinsley 1980]{t80} Tinsley, B. M. 1980, Fund. Cosmic Phys., 5, 287
\bibitem[Tomasko 1970]{t70} Tomasko, M. G. 1970, \apj, 162, 125
\bibitem[Walker 1992]{w92} Walker, A. R. 1992, \pasp, 104, 1063
\bibitem[Willson et al. 1996]{wbs96}Willson, L. A., Bowen, G. H., \& Struck, C. 1996, in From Stars to Galaxies, ed. C. Leitherer, U. Fritze-v. Alvensleben, \& J. Huchra (ASP), 197
\bibitem[Yi 1996]{yi96}  Yi, S. 1996, Ph. D. Thesis, Yale University 
\bibitem[Yi et al. 1995]{yado95}  Yi, S., Afshari, E., Demarque, P., \& Oemler, A. Jr. 1995, \apj, 453, L69
\bibitem[Yi et al. 1997a]{ydk97}  Yi, S., Demarque, P., \& Kim, Y.-C. 1997, \apj, 482, 677 (YDK)
\bibitem[Yi et al. 1997b]{ydo97}  Yi, S., Demarque, P., \& Oemler, A. Jr. 1997, \apj, in preparation
\bibitem[Yi et al. 1993]{yld93}  Yi, S., Lee, Y.-W., \& Demarque, P. 1993, \apj, 411, L25
\end{thebibliography}
\end{document}